\documentclass{aastex62}
\usepackage{amsmath}
\usepackage{bm}
\graphicspath{{./}{figures/}}

\accepted{July 8, 2021}
\submitjournal{ApJ}

\begin{document}

\title{Information-theoretical Limits of Recursive Estimation and Closed-loop
Control in High-contrast Imaging}

\author{Leonid Pogorelyuk}
\affiliation{Department of Aeronautics and Astronautics, Massachusetts Institute of Technology, 77 Massachusetts Avenue, Cambridge, MA 02139, USA}
\author{Laurent Pueyo}
\affiliation{Space Telescope Science Institute, 3800 San Martin Drive Baltimore, MD, US, 21218}
\author{Jared R. Males}
\affiliation{Steward Observatory, University of Arizona, 933 N. Cherry Avenue, Tucson, AZ 85721, USA}
\author{Kerri Cahoy}
\affiliation{Department of Aeronautics and Astronautics, Massachusetts Institute of Technology, 77 Massachusetts Avenue, Cambridge, MA 02139, USA}
\author{N. Jeremy Kasdin}
\affiliation{University of San Francisco, College of Arts and Sciences, 2130 Fulton St., San Francisco, CA, US, 94117}



\begin{abstract}
A lower bound on unbiased estimates of wavefront errors (WFE) is presented for the linear regime of small perturbation and active control of a high-contrast region (dark hole). Analytical approximations and algorithms for computing the closed-loop covariance of the WFE modes are provided for discrete- and continuous-time linear WFE dynamics. Our analysis applies to both image-plane and non-common-path wavefront sensing (WFS) with Poisson-distributed measurements and noise sources (i.e., photon-counting mode). Under this assumption, we show that recursive estimation benefits from infinitesimally short exposure times, is more accurate than batch estimation and, for high-order WFE drift dynamical processes, scales better than batch estimation with amplitude and star brightness. These newly-derived contrast scaling laws are a generalization of previously known theoretical and numerical results for turbulence-driven Adaptive Optics. For space-based coronagraphs, we propose a scheme for combining models of WFE drift, low-order non-common-path WFS (LOWFS) and high-order image-plane WFS (HOWFS) into closed-loop contrast estimates. We also analyze the impact of residual low-order WFE, sensor noise, and other sources incoherent with the star, on closed-loop dark-hole maintenance and the resulting contrast. As an application example, our model suggests that the Roman Space Telescope might operate in a regime that is dominated by incoherent sources rather than WFE drift, where the WFE drift can be actively rejected throughout the observations with residuals significantly dimmer than the incoherent sources. The models proposed in this paper make possible the assessment of the closed-loop contrast of coronagraphs with combined LOWFS and HOWFS capabilities, and thus help estimate WFE stability requirements of future instruments.
\end{abstract}

%\keywords{methods: statistical, data analysis --- techniques: high angular resolution, image processing}


\section{Introduction} 

Wavefront instability is a major limiting factor on the contrast during
coronagraphic observations. In ground-based telescopes, atmospheric
turbulence gives rise to fast wavefront aberrations that are mostly
counteracted by Adaptive Optics (AO) (\cite{roddier1999adaptive}). AO uses 
wavefront sensors and deformable mirrors (DM) to continuously estimate,
predict, and correct the wavefront errors (WFE) using a natural or a laser guide
star. In the Extreme AO regime (smallest wavefront error possible over a small field of view)  the correction precision is fundamentally limited by the photon flux available for wavefront sensing, along with the spatio-temporal properties of atmospheric turbulence (\cite{guyon2005limits,cavarroc2006fundamental}),
resulting in contrasts of about $10^{-6}$  with current 8 m class telescopes (\cite{macintosh2015discovery,currie2018scexao}).

In the absence of atmosphere, space telescopes are expected to achieve
contrasts that are better by at least two orders of magnitude (\cite{demers2015requirements,mennesson2016habitable,bolcar2017large}),
eventually enabling the detection of exo-Earths (\cite{stark2019exoearth,pueyo2019luvoir}).  However, even in pristine environments such as Lagrange L2, space based observatories are subject to small thermal and mechanical disturbances that can result in significant variations of the telescope wavefront and instrument's starlight suppression  (\cite{shaklan2011stability,patterson2015control,perrin2018updated})
% %@INPROCEEDINGS{2018SPIE10698E..09P,
%       author = {{Perrin}, Marshall D. and {Pueyo}, Laurent and {Van Gorkom}, Kyle and {Brooks}, Keira and {Rajan}, Abhijith and {Girard}, Julien and {Lajoie}, Charles-Philippe},
%         title = "{Updated optical modeling of JWST coronagraph performance contrast, stability, and strategies}",
%     booktitle = {Space Telescopes and Instrumentation 2018: Optical, Infrared, and Millimeter Wave},
%          year = 2018,
%       editor = {{Lystrup}, Makenzie and {MacEwen}, Howard A. and {Fazio}, Giovanni G. and {Batalha}, Natalie and {Siegler}, Nicholas and {Tong}, Edward C.},
%       series = {Society of Photo-Optical Instrumentation Engineers (SPIE) Conference Series},
%       volume = {10698},
%         month = aug,
%           eid = {1069809},
%         pages = {1069809},
%           doi = {10.1117/12.2313552},
%       adsurl = {https://ui.adsabs.harvard.edu/abs/2018SPIE10698E..09P},
%       adsnote = {Provided by the SAO/NASA Astrophysics Data System}
% }
As a result, when high contrast imaging of exoplanets is considered, wavefront stability is one of the main drivers of observatories' overall structural designs 
(\cite{coyle2019large})
% @INPROCEEDINGS{2019SPIE11115E..0RC,
%       author = {{Coyle}, Laura E. and {Knight}, J. Scott and {Pueyo}, Laurent and {Arenberg}, Jonathan and {Bluth}, Marcel and {East}, Matthew and {Patton}, Kevin and {Bolcar}, Matthew R.},
%         title = "{Large ultra-stable telescope system study}",
%     booktitle = {UV/Optical/IR Space Telescopes and Instruments: Innovative Technologies and Concepts IX},
%          year = 2019,
%       series = {Society of Photo-Optical Instrumentation Engineers (SPIE) Conference Series},
%       volume = {11115},
%         month = sep,
%           eid = {111150R},
%         pages = {111150R},
%           doi = {10.1117/12.2525396},
%       adsurl = {https://ui.adsabs.harvard.edu/abs/2019SPIE11115E..0RC},
%       adsnote = {Provided by the SAO/NASA Astrophysics Data System}
%
. It also drives the way the data will be collected, e.g., observation scenarios (\cite{bailey2018lessons,laginja2019wavefront}), which in turn can
reduce exoplanet yields due to the overheads necessary to maintain as stable as possible of a wavefront and calibrate remaining variation using post-processing (\cite{stark2019exoearth}).

Increasing the precision of real-time wavefront correction is a topic of much research (\cite{jovanovic2018review,snik2018advances}). Proposed
hardware improvements include faster computers and DMs (\cite{macintosh2018gemini}),
improved architectures of sensors (\cite{ndiaye2013calibration,correia2020performance})
and cameras (\cite{baudoz2005self,bottom2016speckle}),
and use of a space-based laser guide stars (\cite{douglas2019laser}). Algorithmic approaches
aim at exploiting all of the available information for a given system.
They may, for example, utilize the available post-coronagraphic images (\cite{paul2013high,martinache2014sky,miller2017spatial}),
or incorporate WFE dynamics via recursive estimation and predictive
control (\cite{kulcsar2012minimum,males2018predictive,pogorelyuk2019dark}).

Yet, predicting peformances, e.g., computing contrast curves as a function of optical model parameters,
WFE spatio-temporal profiles, control algorithms, and their parameters,
is a complex task. It typically requires running full-model simulations
with various parameter combinations and, possibly, different time
scales. In this work, the authors propose an information theoretical
approach to approximating bounds on the residual WFE given
linear models of their dynamics and of detector sensitivities (at a wavefront sensor and/or image plane). 
The more general treatment of wavefront dynamics presented here allows for the
examination of both batch and recursive estimation and both common and non-common
path control loops.

\begin{table}
\caption{Table of symbols}
\begin{tabular}{|l|c|c|l|}
\hline 
symbol & units & domain & definition\tabularnewline
\hline 
$t_{s}$ & $\mathrm{s}$ & $\mathbb{R}$ & Sampling time\tabularnewline
\hline 
$\bm{\epsilon}$ & $\mathrm{nm}$ & $\mathbb{R}^{r}$ & Controllable wavefront modes\tabularnewline
\hline 
$r$ & N/A & $\mathbb{N}$ & Number of wavefront modes\tabularnewline
\hline 
$\dot{N}_{S}$ & $\mathrm{s}^{-1}$ & $\mathbb{R}$ & Photon flux from star at the primary mirror\tabularnewline
\hline 
$G_{i}$ & $\mathrm{nm}^{-1}$ & $\mathbb{R}^{2c\times r}$ & Sensitivity of electric field to wavefront modes\tabularnewline
\hline 
$c$ & N/A & $\mathbb{N}$ & Number of wavelengths incoherently summed\tabularnewline
\hline 
$\mathbf{E}_{0,i}$ & $1$ & $\mathbb{R}^{2c}$ & Static (and uncontrollable) component of the E-field\tabularnewline
\hline 
$I_{i}=\dot{N}_{S}\left\Vert G_{i}\mathbf{\bm{\epsilon}}+\mathbf{E}_{0,i}\right\Vert ^{2}$ & $\mathrm{s}^{-1}$ & $\mathbb{R}$ & Photon flux at pixel $i$\tabularnewline
\hline 
$D_{i}$ & $\mathrm{s}^{-1}$ & $\mathbb{R}$ & Flux from sources incoherent with speckles\tabularnewline
\hline 
$y_{i}\sim poisson\left(\left(I_{i}+D_{i}\right)t_{s}\right)$ & 1 & $\mathbb{N}$ & Measured number of photons at pixel $i$\tabularnewline
\hline 
$\Lambda_{j}^{2}=\underset{i,l}{\sum}\left[G_{i}\right]_{lj}^{2}$ & $\mathrm{nm}^{-2}$ & $\mathbb{R}$ & Sensitivity to mode $j$\tabularnewline
\hline 
$P,p^{2}$ & $\mathrm{nm}^{2}$ & $\mathbb{R}^{r\times r},\mathbb{R}^{1}$ & Wavefront estimate error covariance\tabularnewline
\hline 
$Q$ & $\mathrm{nm}^{2}$ & $\mathbb{R}^{r\times r}$ & Wavefront drift covariance\tabularnewline
\hline 
${\cal I}$ & $\mathrm{nm}^{-2}$ & $\mathbb{R}^{r\times r}$ & Fisher information that all $y_{i}$ carry about $\mathbf{\bm{\epsilon}}$\tabularnewline
\hline 
$\Xi,\xi^{2}$ & $\mathrm{nm}^{2}\mathrm{s}^{-1}$ & $\mathbb{R}^{r\times r},\mathbb{R}^{1}$ & Wavefront drift diffusion matrix/coefficient\tabularnewline
\hline 
$\cdot{}^{WS}$ & N/A & N/A & Superscript for wavefront sensing quantities\tabularnewline
\hline 
$\cdot{}^{IP}$ & N/A & N/A & Superscript for image plane quantities\tabularnewline
\hline 
$C=\dot{N}_{S}^{-1}\sum\left(I_{i}^{IP}+D_{i}^{IP}\right)$ & $1$ & $\mathbb{R}$ & Average contrast\tabularnewline
\hline 
$C_{0}=\sum\left\Vert \mathbf{E}_{0,i}^{IP}\right\Vert ^{2}=\left\Vert \mathbf{E}_{0}^{IP}\right\Vert ^{2}$ & $1$ & $\mathbb{R}$ & Average raw contrast\tabularnewline
\hline 
$f,f_0$ & $\mathrm{s}^{-1}$ & $\mathbb{R}$ & Frequency, knee frequency\tabularnewline
\hline 
$\theta^{2}$ & $\mathrm{nm}^{2}\mathrm{s}^{1}$ & $\mathbb{R}$ & Wavefront PSD in the limit $f=0$\tabularnewline
\hline 
$\gamma$ & N/A & $\mathbb{N}$ or $\mathbb{R}$ & Order of WFE dynamics \tabularnewline
\hline 
$u$ & $\mathrm{nm}\mathrm{s}^{-1}$ & $\mathbb{R}$ & Wavefront drift rate\tabularnewline
\hline 
$v,w$ & $\mathrm{s}^{-\frac{1}{2}}$ & $\mathbb{R}$ & Continuous-time white noise\tabularnewline
\hline 
\end{tabular}
\end{table}

In Section~\ref{sec:bounds} we outline a technique for computing
a bound on the variance of the WFE estimates based on the Cram{\'e}r-Rao
inequality (\cite{rao1945information,cramer1946contribution}). The discrete-time
and continuous-time versions of the variance bounds can then be used
to estimate the residual starlight intensity in the image plane (or
contrast). In Section~\ref{sec:special_cases}, closed-form expressions
for the contrast are derived for some special cases. In particular,
we provide scaling laws for the WFE variance as a function of 
drift magnitude, power spectral density (PSD), star brightness, and detector noise. The newly derived scaling laws are then compared to those of existing AO systems, and find broad overall agreement.
Section~\ref{sec:application_examples} contains application examples
in the context of space-based coronagraphs. It discusses the connection
between low- and high-order wavefront sensing (LOWFS and HOWFS), and
presents HOWFS closed-loop bounds for the Nancy Grace Roman Space
Telescope (RST). Section~\ref{sec:conclusions} summarizes the work.

\section{\label{sec:bounds}Approximate Bounds of Unbiased WFE Mode Estimates}

Throughout the paper we assume that the telescope operates in a steady-state
linear regime after achieving its best contrast. In the case of
the RST, for example, our analysis does not apply to dark hole creation (\cite{krist2015numerical})
via pair-probing and EFC (\cite{give2011pair}). Instead, the focus of
Section~\ref{sub:discrete_time} is on the slow ``drift'' of wavefront 
aberrations during the long scientific observation (tens of hours)
of a relatively dim target. 

We work under the assumption that a nominal dark hole has been generated using the methods above. When seeking to maintain this dark hole in the presence thermal or mechanical drifts, such as in the optical tube assembly (OTA), the  information ``about'' WFE modes (and hence the ability to correct them), diminishes as the time since they were last estimated increases.  However, the information contained in each wavefront sensing measurement that can be used to correct the WFE estimates increases as a function of exposure time. 
Formulating this information balance allows, under certain assumptions,
the estimation of a bound on the residual (closed-loop) WFE and, hence, the
contrast.

Throughout the discussion, the $r$ WFE modes \textit{coefficients} will be denoted as $\bm{\epsilon}$ (before correction, or open-loop) and $\bm{\epsilon}^{CL}$ (closed-loop). The closed-loop contrast depends on ``how far on average'' $\bm{\epsilon}^{CL}$ is from zero (its temporal covariance) and the sensitivity of the image-plane speckles to $\bm{\epsilon}^{CL}$, denoted by $G^{IP}$. The challenging part, however, is determining the covariance of $\bm{\epsilon}^{CL}$ that ties directly to contrast, without full end-to-end simulations of the closed-loop wavefront sensor and DM operations. Indeed, even under the assumption of a perfect controller,  closed-loop wavefront covariance still depends on open loop wavefront properties, wavefront sensor architecture, reconstruction algorithm, incident flux and detector properties. In this paper we present a theoretical framework that captures all these parameters while circumventing the need for full closed-loop simulations.

%as it depends on the measurements which themselves depend on $\bm{\epsilon}^{CL}$ in a probabilistic way. Moreover, an ``optimal'' wavefront sensor would utilize all previous measurements and weight them according to ``how fast'' the open-loop WFE coefficients, $\bm{\epsilon}$, drift.

\begin{figure}
\includegraphics{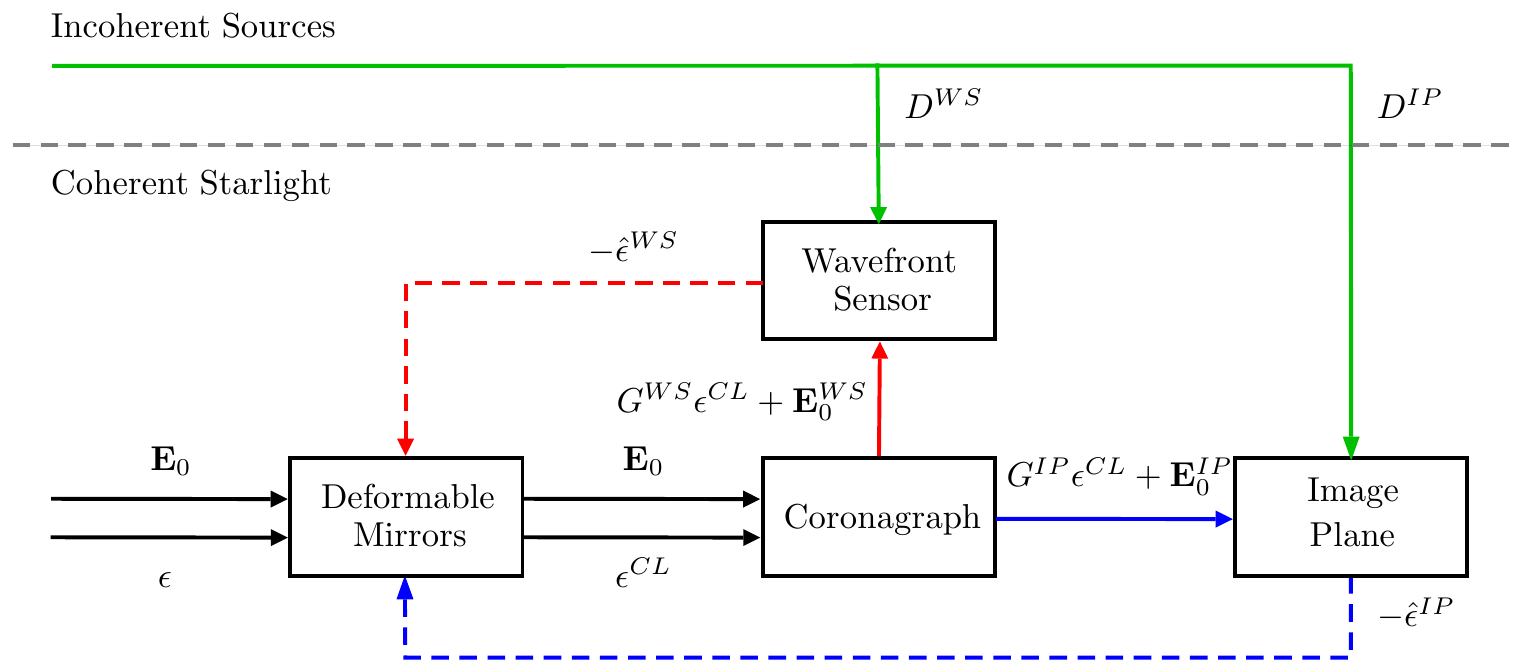}\caption{\label{fig:diagram_simplified} A schematic of a coronagraph with wavefront estimation from measurements taken at both a dedicated sensor (denoted by the superscript $\cdot^{WS}$) and at the science camera, i.e., at the image plane (denoted by $\cdot^{IP}$). The open-loop coefficients of WFE modes, $\bm{\epsilon}$, are partially corrected by the DM based on their estimates, $\hat{\bm{\epsilon}}^{WS}$ and $\hat{\bm{\epsilon}}^{IP}$, resulting in the closed-loop modes, $\bm{\epsilon}^{CL}$. Under linearity assumptions typical in high-contrast imaging, the electric fields of the starlight speckles at the wavefront sensor and the image plane depend linearly on the WFE modes through their sensitivity matrices, $G^{WS}$ and $G^{IP}$, respectively. The static component of the electric field (not affected by the DM), is denoted by $\mathbf{E}_0^{WS/IP}$, and sources incoherent with the starlight are denoted by $D^{WS/IP}$ (although not implicitly shown, some of the incoherent sources also pass through the coronagraph). Multiplying the squared magnitude of the electric field by the photon flux at the primary mirror, $\dot{N}_S$, gives the flux at the sensor/image (discretized by pixels and wavelengths). In the examples discussed in this paper, we consider both the cases of a wavefront sensor distinct from image plane, and using the latter to compute DM feedback (with wavefront sensor residuals contributing to the incoherent sources at the image plane, see Sec. \ref{sub:combining_bounds}). }
\end{figure}

Note that, as depicted in Fig.~\ref{fig:diagram_simplified}, the wavefront sensing can be performed either in the image plane, denoted by the superscript $\cdot^{IP}$, or at a dedicated non common-path wavefront sensor, denoted by $\cdot^{WS}$. When the analysis is applicable to both cases, we drop the superscript (i.e., we use a general electric field WFE sensitivity matrix $G$ instead of $G^{IP}$ or $G^{WS}$). Additional sensor quantities that affect estimation include the static electric field, $\mathbf{E}_0$ (the constant-in-time zeroth order term in the expansion of the electric field in terms of $\bm{\epsilon}$), and photon sources that are not affected by control, $D$ (e.g., dark current and post-LOWFS residuals in the image plane, see Sec.~\ref{sub:combining_bounds}).

In Section~\ref{sub:discrete_time}, we focus on the ``simplest'' drift scenario, under a discrete time approximation, for which we derive an implicit equation that relates open and closed-loop WFE modes covariance per iteration of the WFS system. The impact of each WFE mode on the contrast  is assumed to be proportional to its closed-loop covariance. This formulation also assumes an open loop temporal power spectral density inversely proportional to frequency, scaling as $1/f$. In Section~\ref{sub:continuous_time}, we extend our theoretical bound to any generic open loop temporal PSD, using a continuous time formulation that is more suitable for computing bounds of AO, LOWFS, and non-common path HOWFS. The numerical algorithms for computing these bounds are given in Section~\ref{sub:implementation}.

\subsection{\label{sub:discrete_time}Derivation of Bounds for Brownian Motion WFE Drift (Discrete-time Formulation)}

Below, we first describe our assumptions for the open- and closed-loop WFE coefficients, $\bm{\epsilon}_{k}$ and $\bm{\epsilon}^{CL}_{k}$, where $k$ denotes the number of the exposure. We then relate $\bm{\epsilon}^{CL}_{k}$ to the number of photons detected during the $t_s$ long exposures, and the Fisher information ${\cal I}_{k}$ contained within those wavefront sensing measurements (either using the science camera or a dedicated sensor) about $\bm{\epsilon}^{CL}_{k}$. This relationship allows us to use the Cram{\'e}r-Rao inequality to incorporate the uncertainties due to both WFE drift and shot-noise into a single implicit equation from which the covariance of $\bm{\epsilon}^{CL}_{k}$ (denoted by $P_{k}$) can be estimated. Finally, the estimate of the ``average'' $P_k$ in steady state ($k\rightarrow\infty$) is used to get the bounds on the closed-loop contrast.

\subsubsection{WFE modes drift}
We begin with a simple Brownian Motion (\cite{durrett2019probability})
model for the evolution of high-order WFE modes in the context of space based coronagraphs. This assumption leads to linear growth of uncertainty in the intensity -- an approximation that is commonly used when evaluating exoplanet detection performance (\cite{nemati2020method}).
%based on the numerical analysis of Observing Scenario (OS) 9 of RST %(\cite{krist2020observing}) (see sec.~\ref{sub:RST}). 
Formally, the $r$ WFE mode coefficients, $\bm{\epsilon}_{k}=\bm{\epsilon}(k\cdot t_{s})\in\mathbb{R}^{r}$,
are such that their increments are normally (and independently) distributed with some drift covariance $Q\in\mathbb{R}^{r\times r}$,

\[
\bm{\epsilon}_{k+1}-\bm{\epsilon}_{k}\sim{\cal N}\left(\mathbf{0},Q\right),\: Q(t_{s})>0.
\]
We make the additional simplifying assumption that there exists an
\emph{unbiased} estimate of WFE modes, $\hat{\mathbf{\bm{\epsilon}}}_{k}$,
whose error is also normally distributed with covariance $P_{k}$,

\[
\hat{\mathbf{\bm{\epsilon}}}_{k}-\bm{\epsilon}_{k}\sim{\cal N}\left(\mathbf{0},P_{k}\right),\:P_{k}>0,
\]
(independently of the WFE increments). Furthermore, the DMs are assumed
to be able to perfectly reproduce the WFE modes. Although, due to the
imperfect knowledge of these modes and the inability of the estimator to predict their increments, the corrections are slightly off.
We call them the ``closed-loop'' WFE modes,

\[
\bm{\epsilon}_{k+1}^{CL}=\bm{\epsilon}_{k+1}-\hat{\mathbf{\bm{\epsilon}}}_{k},
\]
and they are, too, normally distributed with, 
\begin{equation}
\bm{\epsilon}_{k+1}^{CL}\sim{\cal N}\left(\mathbf{0},P_{k}+Q\right).\label{eq:normal_assumption}
\end{equation}
Note that $Q$ may now also ``contain'' actuator drift, i.e., the wavefront changes faster if each DM actuator also exhibits Brownian motion on top of the prescribed commands (more complex DM dynamics can be treated in a manner suggested in Sec.~\ref{sub:continuous_time}).

\subsubsection{Measurements model and Fisher information}
Here, we relate the closed-loop WFE modes $\bm{\epsilon}_{k}^{CL}$ to the probabilistic photon measurements. Our goal is to find an expression for the information ${\cal I}$ that the measurements carry about the modes, to be later used in the Cram{\'e}r-Rao inequality.
The discussion is constrained to the linear regime where the sensitivity of the field to the WFE modes at detector pixel $i$ is $G_{i}\in\mathbb{R}^{2c\times r}$ and $c$ is
the number of wavelengths in the spectral discretization in the model. Together with the static (and presumably known) component, $\mathbf{E}_{0,i}\in\mathbb{R}^{2c}$, the electric field at pixel $i$ is given by $G_{i}\mathbf{\bm{\epsilon}}^{CL}+\mathbf{E}_{0,i}\in\mathbb{R}^{2c}$.

The fields are scaled such that the photon arrival rate (intensity)
at pixel $i$ is given by
\[
I_{i}=\dot{N}_{S}\left\Vert G_{i}\mathbf{\bm{\epsilon}}^{CL}+\mathbf{E}_{0,i}\right\Vert ^{2},
\]
where $\dot{N}_{S}$ is the photon flux from the star integrated over the primary mirror of the telescope and propagated thorough the various optics (reflective surfaces, masks) between the primary and the wavefront sensing detector. Photon flux from external sources such as zodiacal dust, $D_{i}^{ext}$,
is presumably fixed and known (or, at least, its average contribution
can be canceled out by image subtraction). The flux of internal sources
of photoelectrons such as clock-induced charge and dark current (\cite{harding2015technology})
is denoted as $D_{i}^{int}$ and is also assumed to be known.

The probability distribution of the measured number of photons, $y_{i}$, in photon-counting
mode can be a complex function of $I_{i},D_{i}^{ext},D_{i}^{int}$
and sampling time $t_{s}$ (\cite{hirsch2013stochastic,hu2020sequential}).
Here, we assume it is the Poisson distribution,
\begin{equation}
\mathrm{pmf}(y_{i})=\frac{1}{y_{i}!}\left(\left(I_{i}+D_{i}\right)t_{s}\right)^{y_{i}}e^{-\left(I_{i}+D_{i}\right)t_{s}},\label{eq:poisson}
\end{equation}
with $D_{i}=D_{i}^{ext}+D_{i}^{int}$, although the analysis below
can be repeated with any probability mass function, $\mathrm{pmf}(y_{i})$. 
This assumption holds well if continuously-distributed noise sources (such as clock-induced charge) are small enough (\cite{wilkins2014characterization}) as to not cause confusion with the number of detected photons. Besides simplifying the discussion, it is justifiable in the context of finding lower bounds on contrast, and leads to a conclusion that shorter exposure times are always preferable (see Sec.~\ref{sub:special_discrete}).

The Fisher
information that the measured number of photons, $\left\{ y_{i}\right\} $,
carry about the WFE modes, $\mathbf{\bm{\epsilon}}^{CL}$, is given
by
\[
{\cal I}=\underset{i}{\sum}\mathrm{E}_{y_{i}}\left\{ \left(\frac{\partial\log \mathrm{pmf}(y_{i})}{\partial\bm{\epsilon}^{CL}}\right)\left(\frac{\partial\log \mathrm{pmf}(y_{i})}{\partial\bm{\epsilon}^{CL}}\right)^{T}\right\} \in\mathbb{R}^{r\times r},
\]
where $\mathrm{E}_{y_{i}}\left\{ \cdot\right\} $ denotes the expectation
w.r.t. $y_{i}$. In particular, with $\mathrm{pmf}(y_{i})$ given by Eq.~(\ref{eq:poisson}),

\begin{equation}
{\cal I}=\underset{i}{\sum}\frac{4\dot{N}_{S}t_{s}}{\left\Vert G_{i}\mathbf{\bm{\epsilon}}^{CL}+\mathbf{E}_{0,i}\right\Vert ^{2}+\dot{N}_{S}^{-1}D_{i}}G_{i}^{T}\left(G_{i}\mathbf{\bm{\epsilon}}^{CL}+\mathbf{E}_{0,i}\right)\left(G_{i}\mathbf{\bm{\epsilon}}^{CL}+\mathbf{E}_{0,i}\right)^{T}G_{i}.\label{eq:information}
\end{equation}
This information can be used to compute an estimate of
the WFE modes based on a single sensing iteration (such methods will be referred to as ``batch estimation''), or combined with the information contained in previous estimates (i.e., ``recursive estimation'').

\subsubsection{An implicit equation to estimate the covariance of the closed-loop WFE modes}

We are now ready to combine the probabilistic assumptions about the evolution of the WFE modes leading to Eq.~(\ref{eq:normal_assumption}) with the measurement model that gives Eq.~(\ref{eq:information}), to get an equation from which a bound on the WFE modes covariance $P$ can be estimated. From Eq.~(\ref{eq:normal_assumption}),
the Fisher information that the estimate, $\hat{\mathbf{\bm{\epsilon}}}_{k}$,
carries about the modes, $\bm{\epsilon}_{k+1}^{CL}$, is $\left(P_{k}+Q\right)^{-1}$. 
%Therefore the information contained in both the new measurements and the previous estimate is ${\cal I}_{k+1}+\left(P_{k}+Q\right)^{-1}$.
Together with the information contained in the new measurements, ${\cal I}_{k+1}$, the information about the new estimate ($\hat{\mathbf{\bm{\epsilon}}}_{k+1}$) is therefore ${\cal I}_{k+1}+\left(P_{k}+Q\right)^{-1}$.

First, we apply the Cram{\'e}r-Rao inequality (\cite{rao1945information,cramer1946contribution})
which states that the variance $P_{k+1}$ of the unbiased (recursive) estimate,
$\hat{\mathbf{\bm{\epsilon}}}_{k+1}$, is greater than the reciprocal
of the Fisher information,
\[
P_{k+1}\ge\left({\cal I}_{k+1}+\left(P_{k}+Q\right)^{-1}\right)^{-1}.
\]
This inequality captures the fundamental trade-off associated with closed loop WFS or AO operations: the  information about the open-loop drift obtained during a sensing exposure, ${\cal I}_{k+1}$, competes with the accrued open-loop variance during that exposure, $Q$. The closed-loop variance at iteration $k+1$ cannot be smaller than the combination of these two phenomena. Since we are interested in an estimate of the covariance, $P$, of the residual
WFE modes in steady-state operation, we assume that it doesn't change
much ($P_{k+1}\approx P_{k}\approx P$). It is therefore reasonable to approximate the Fisher information with a constant that is equal to the average of Eq.~(\ref{eq:information}) across ``all'' exposures (expectation), 
\[
{\cal I}_{k+1}\approx{\cal I}_{k}\approx\mathrm{E}_{\bm{\epsilon}^{CL}}\left\{ \left.{\cal I}\right|P+Q\right\} ,
\]
where $\mathrm{E}_{\bm{\epsilon}^{CL}}\left\{ \left.\cdot\right|P+Q\right\} $
denotes expectation w.r.t. $\bm{\epsilon}^{CL}\sim{\cal N}\left(\mathbf{0},P+Q\right)$. (Here we implicitly assumed that the WFE remain constant throughout the exposure; A more complete analysis is given in Appendix~\ref{sec:finite_exposure} and leads to qualitatively identical conclusions.)

Finally, in order to estimate a lower bound on $P$, we replace the Cram{\'e}r-Rao inequality with an equality and solve it in a slightly modified form,
\begin{equation}
P^{-1}-\left(P+Q\right)^{-1}=\mathrm{E}_{\bm{\epsilon}^{CL}}\left\{ \left.{\cal I}\right|P+Q\right\} .\label{eq:bound}
\end{equation}
Note that the averaging in the above equation makes it independent of the time varying $\bm{\epsilon}^{CL}$ and therefore self-contained, although it also makes finding a solution more challenging as discussed in Sec.~\ref{sub:implementation}.
For batch estimation (when all information contained in previous estimates is discarded), the bound can be found by solving

\begin{equation}
P_{batch}^{-1}=\mathrm{E}_{\bm{\epsilon}^{CL}}\left\{ \left.{\cal I}\right|P_{batch}+Q\right\} ,\label{eq:batch_bound}
\end{equation}
instead. 

\subsubsection{An expression for the closed-loop contrast}

Equipped with an estimate of the steady-state residual WFE covariance
$P+Q$ (calculated in Sec.~\ref{sub:implementation} based on Eq.~(\ref{eq:bound})), we wish to find the average contrast across the image plane, $C$. 

First, the intensity at the image plane (denoted by $\cdot^{IP}$) at pixel $i$ can be averaged w.r.t. the WFE modes $\bm{\epsilon}^{CL}$,
\[
\mathrm{E}_{\bm{\epsilon}^{CL}}\left\{ \left.I_{i}^{IP}\right|P+Q\right\}  =\dot{N}_{S}^{IP}\cdot\left(\mathrm{trace}\left\{ G_{i}^{IP}\left(P+Q\right)\left(G_{i}^{IP}\right)^{T}\right\} +\left\Vert \mathbf{E}_{0,i}^{IP}\right\Vert ^{2}\right),
\]
which follows directly from Eq.~(\ref{eq:normal_assumption}) and the definitions of $I_i$ and $\mathrm{E}_{\bm{\epsilon}^{CL}}$ (the cross term $\mathrm{E}_{\bm{\epsilon}^{CL}}\left\{ \left.\mathbf{E}_{0,i}^{IP}\left(G_{i}^{IP}\mathbf{\bm{\epsilon}}^{CL}\right)^{T}\right|P+Q\right\}$ is zero because $\mathbf{\bm{\epsilon}}^{CL}$ is zero-mean and $\mathbf{E}_{0,i}$ is constant). Note that $\dot{N}_{S}^{IP}$ now refers to the star's photon flux at the primary mirror, but only in the bandwidth of the sensors at the image plane detector.

While the time-averaged pixel-wise intensity can be used to compute contrast curves, it is also useful to have a single scalar that describes the closed-loop performance of the coronagraph. To this end, we define the average contrast as the sum of all intensities (except exoplanets) across all image plane pixels, normalized by the photon flux from the star at the primary mirror (in the bandwidth of the image-plane detectors),
\[
C=\frac{\underset{i}{\sum}\left(\mathrm{E}_{\bm{\epsilon}^{CL}}\left\{ \left.I_{i}^{IP}\right|P+Q\right\}+D_{i}^{IP}\right)}{\dot{N}_{S}^{IP}} .
\]
In terms of the WFE covariance, the (average) contrast is given by
\begin{alignat}{1}
C & =C_{0}+\underset{i}{\sum}\left[\frac{D_{i}^{IP}}{\dot{N}_{S}^{IP}}+\mathrm{trace}\left\{ G_{i}^{IP}\left(P+Q\right)\left(G_{i}^{IP}\right)^{T}\right\} \right].\label{eq:contrast}
\end{alignat}
where $C_{0}=\underset{i}{\sum}\left\Vert \mathbf{E}_{0,i}^{IP}\right\Vert ^{2}$
is the (average) raw contrast in the absence of WFE and incoherent sources. Note that the error in the speckle's contribution to the contrast is directly proportional to the error in the covariance estimate $P+Q$.

\subsection{\label{sub:continuous_time}Continuous-time Formulation}

The discussion in Sec.~\ref{sub:discrete_time} can be repeated with more general dynamical models
for WFE drift suitable for finding bounds on contrasts of ground-based telescopes. Below, this is illustrated with a continuous-time
system that may approximate a larger family of temporal power spectral
densities (PSD) that are typical of AO (see Sec.~\ref{sub:special_AO}).
We extend the analysis to include the time derivatives of the WFE modes, $\frac{d}{dt}\bm{\epsilon},\frac{d^{2}}{dt^{2}}\bm{\epsilon},....$, and their estimates, $\widehat{\frac{d}{dt}\bm{\epsilon}},\widehat{\frac{d^{2}}{dt^{2}}\bm{\epsilon}},...$. Our goal is to get their closed-loop covariances, i.e., a higher-order continuous-time equivalent of Eq.~(\ref{eq:bound}) from which they can be found.

We assume that the dynamics of the open-loop WFE modes, $\bm{\epsilon}$,
are linear and given by a $\gamma$-th order transfer function between white noise, $\mathbf{v}$, and $\bm{\epsilon}$. This can be stated as
\begin{equation}
\begin{bmatrix}\frac{d}{dt}\bm{\epsilon}\\
\vdots\\
\frac{d^{\gamma}}{dt^{\gamma}}\bm{\epsilon}
\end{bmatrix}=A\begin{bmatrix}\bm{\epsilon}\\
\vdots\\
\frac{d^{\gamma-1}}{dt^{\gamma-1}}\bm{\epsilon}
\end{bmatrix}+B\mathbf{v}(t),\label{eq:continous}
\end{equation}
where $\mathbf{v}(t)\in\mathbb{R}^{s}$, and $A\in\mathbb{R}^{\gamma r\times\gamma r}$
and $B\in\mathbb{R}^{\gamma r\times s}$ are some matrices describing the temporal evolution of the WFE and how it is forced by the white noise.
Instead of the discrete-time error covariance $P+Q$ of the WFE modes estimate, a continuous-time covariance $\Pi_{1,1,}$ will be used. However, one has to keep in mind the uncertainties in the estimates of the derivatives of the WFE modes as well. We assume that
 the full state estimate (including time derivatives) is normally distributed with covariance $\Pi$, i.e.,
\[
\begin{bmatrix}\hat{\bm{\epsilon}}\\
\vdots\\
\widehat{\frac{d^{\gamma-1}}{dt^{\gamma-1}}\bm{\epsilon}}
\end{bmatrix}\sim{\cal N}\left(\mathbf{0},\Pi\right),
\]
The $\Pi_{1,1}\in\mathbb{R}^{r\times r}$ matrix is then a sub-matrix of $\Pi\in\mathbb{R}^{\gamma r\times \gamma r}$ appearing first on its main diagonal.
The steady-state information rate is given by dividing Eq.~(\ref{eq:information})
by $t_{s}$ and taking the expectation w.r.t. $\bm{\epsilon}^{CL}\sim{\cal N}\left(\mathbf{0},\Pi_{1,1}\right)$,

\[
\dot{{\cal I}}(\Pi_{1,1})=\mathrm{E}_{\bm{\epsilon}^{CL}}\left\{ \left.\underset{i}{\sum}\frac{4\dot{N}_{S}}{\left\Vert G_{i}\mathbf{\bm{\epsilon}}^{CL}+\mathbf{E}_{0,i}\right\Vert ^{2}+\dot{N}_{S}^{-1}D_{i}}G_{i}^{T}\left(G_{i}\mathbf{\bm{\epsilon}}^{CL}+\mathbf{E}_{0,i}\right)\left(G_{i}\mathbf{\bm{\epsilon}}^{CL}+\mathbf{E}_{0,i}\right)^{T}G_{i}\right|\Pi_{1,1}\right\} .
\]

We then derive the continuous version of Eq.~(\ref{eq:bound}), where we use $\Pi$ instead of $P$. Here we do not provide details and instead direct the reader to the derivation of the Kalman-Bucy filter (see, for example,
\cite{stengel1994optimal}). In steady state, $\Pi$ is the solution
of

\begin{equation}
0=A\Pi+\Pi A^{T}+BB^{T}-\Pi\begin{bmatrix}\dot{{\cal I}}(\Pi_{1,1}) & 0 & \cdots\\
0 & 0\\
\vdots &  & \ddots
\end{bmatrix}\Pi,\label{eq:Pi_equation}
\end{equation}
which needs to be solved instead of Eq.~(\ref{eq:bound}) in the continuous time case.
The contrast is then given by
\[
C=C_{0}+\underset{i}{\sum}\left[\frac{D_{i}^{IP}}{\dot{N}_{S}^{IP}}+\mathrm{trace}\left\{ G_{i}^{IP}\Pi_{1,1}\left(G_{i}^{IP}\right)^{T}\right\} \right].
\]

\subsection{\label{sub:implementation}Implementation}
Knowing the parameters of the linearized system (WFE sensitivity $G$, static field $\mathbf{E}_0$, fluxes $\dot{N}_S$ and $D$, and drift covariance $Q$), is sufficient to find the closed-loop WFE covariance, $P$, via Eq.~(\ref{eq:bound}) (or, Eq.~(\ref{eq:Pi_equation}) in the continuous time case where $A$ and $B$ describe the dynamics instead of $Q$). However, the equation is challenging to solve as is, since it involves the expectation $\mathrm{E}_{\bm{\epsilon}^{CL}}$ --  an integral which depends non-linearly on the unknown matrix, $P$. Instead, we propose a random-sampling and an analytical-approximation approach for the discrete-time and continuous-time cases, respectively.

\subsubsection{A random-sampling approach to finding the discrete-time WFE covariance}

We introduce an iterative algorithm for computing the steady-state WFE covariance estimates, $P$, based on Eq.~(\ref{eq:bound}). Instead of explicitly computing $\mathrm{E}_{\bm{\epsilon}^{CL}}$, the algorithm samples the WFE coefficients, $\bm{\epsilon}^{CL}_{k+1}$, given the covariance $P_k$, and uses them to compute the Fisher information, ${\cal I}_{k+1}$ which is then used to compute $P_{k+1}$.

\paragraph*{Algorithm 1 - Discrete Time (Brownian Motion)}
\begin{enumerate}
\item Initialize $P_{k=0}$ 
\item Sample $\mathbf{\bm{\epsilon}}_{k+1}^{CL}\sim{\cal N}\left(\mathbf{0},P_{k}+Q\right)$
\item Compute ${\cal I}_{k+1}$ via Eq.~(\ref{eq:information})
\item Advance via $P_{k+1}=\left(\left(P_{k}+Q\right)^{-1}+{\cal I}_{k+1}\right)^{-1}$ (for batch estimation, via $P_{k+1}={\cal I}_{k+1}^{-1}$)
\item Repeat steps 2 to 4 until the average of the covariance estimate $P=\frac{1}{k+1}\underset{l=1}{\overset{k+1}{\sum}}P_{l}$
has converged ($\frac{1}{k}\left\Vert P_{k+2}-\frac{1}{k+1}\underset{l=1}{\overset{k+1}{\sum}}P_{l}\right\Vert $ remains arbitrarily small)
\end{enumerate}

Note that the covariances $P_k$ depend on the randomly sampled $\bm{\epsilon}^{CL}_{k}$ and are therefore also random, although \textit{their average} tends to converge to $P$ -- the final covariance estimate. After computing $P$, the contrast can be found via Eq.~(\ref{eq:contrast}), in which $G^{IP}$ stands for the image-plane sensitivity.

\subsubsection{An analytical approximation of the Fisher information}

\begin{figure}
\includegraphics[scale=0.75]{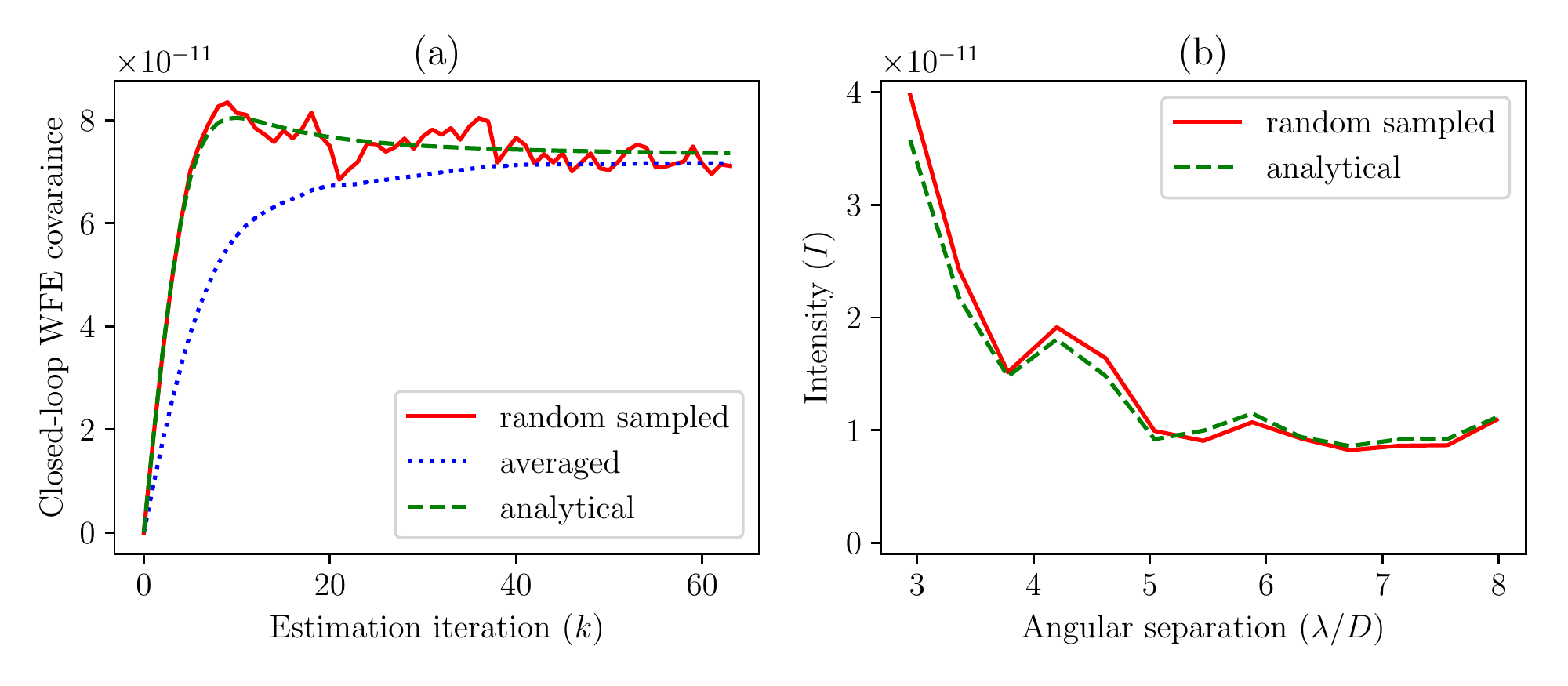}\caption{\label{fig:convergence}Illustration of the convergence of Algorithm~1 based on RST OS 9 data (for details about the data, the dimensionless scaling and computation of the $G$ matrix see Sec.~\ref{sub:RST}) (a) Covariance of first mode (the top left element of $P_k$)
as a function of iterations of the algorithm (solid
red line) and its average (dotted blue line). When using Eq.~(\ref{eq:EI_approx})
to compute the information matrix, the covariance converges to a similar
value (dashed green line). (b) The steady-state intensity averaged across the dark hole as a function of angular separation
computed based on a randomly sampled WFE (per Algorithm~1,
solid red line) and based on Eq.~(\ref{eq:EI_approx}) (dashed green line).}
\end{figure}

Instead of random sampling, based on Eq.~(\ref{eq:information}), one may approximate the expected information, $\mathrm{E}_{\bm{\epsilon}^{CL}}\left\{ \left.{\cal I}\right|P_{k+1}+Q\right\} $, to get a smooth convergence
at the expense of some precision. This is achieved by replacing $\left\Vert G_{i}\mathbf{\bm{\epsilon}}^{CL}+\mathbf{E}_{0,i}\right\Vert ^{2}$ and $\left(G_{i}\mathbf{\bm{\epsilon}}^{CL}+\mathbf{E}_{0,i}\right)\left(G_{i}\mathbf{\bm{\epsilon}}^{CL}+\mathbf{E}_{0,i}\right)^{T}$ by their expectation,
\begin{equation}
\mathrm{E}_{\bm{\epsilon}^{CL}}\left\{ \left.{\cal I}\right|P+Q\right\} \approx\underset{i}{\sum}\frac{4\dot{N}_{S}t_{s}}{\mathrm{trace}\left\{ G_{i}\left(P+Q\right)G_{i}^{T}+\mathbf{E}_{0,i}\mathbf{E}_{0,i}^{T}\right\} +\dot{N}_{S}^{-1}D_{i}}G_{i}^{T}\left(G_{i}\left(P+Q\right)G_{i}^{T}+\mathbf{E}_{0,i}\mathbf{E}_{0,i}^{T}\right)G_{i}.\label{eq:EI_approx}
\end{equation}
Figure~\ref{fig:convergence} illustrates the difference between
computing $P$ based on random sampling of WFE modes (step 3 of Algorithm 1) and
the above analytical approximation. The latter is almost as precise,
and will be used in the analysis in the next section.

\subsubsection{An analytical-approximation approach to finding the continous-time WFE covariance}
In order to solve Eq.~(\ref{eq:Pi_equation}), one has to propagate the full-state covariance matrix, $\Pi$, in continuous time. The algorithm below does that by first approximating the information rate via Eq.~(\ref{eq:EI_approx}), and then updating the estimate of $\Pi$ via the forward Euler method. The time-step $\Delta t$ needs to be small enough so that $\Pi$ does not diverge, and we suspect that more sophisticated numerical schemes can result in faster convergence.

\paragraph*{Algorithm 2 - Continuous Time (Arbitrary Linear Dynamics)}
\begin{enumerate}
\item Initialize $\Pi(t=0)$ and pick $\Delta t$
\item Compute $\dot{{\cal I}}(t)$ via
\[
\dot{{\cal I}}(t)=\underset{i}{\sum}\frac{4\dot{N}_{S}}{\mathrm{trace}\left\{ G_{i}\Pi_{1,1}(t)G_{i}^{T}+\mathbf{E}_{0,i}\mathbf{E}_{0,i}^{T}\right\} +\dot{N}_{S}^{-1}D_{i}}G_{i}^{T}\left(G_{i}\Pi_{1,1}(t)G_{i}^{T}+\mathbf{E}_{0,i}\mathbf{E}_{0,i}^{T}\right)G_{i}
\]

\item Advance via
\[
\Pi(t+\Delta t)=\Pi(t)+\left(A\Pi+\Pi A^{T}+BB^{T}-\Pi\begin{bmatrix}\dot{{\cal I}}(t) & 0 & \cdots\\
0 & 0\\
\vdots &  & \ddots
\end{bmatrix}\Pi\right)\Delta t
\]

\item Repeat steps 2 and 3 until $\Pi$ has converged
\end{enumerate}

Note that since we used an analytical approximation of the Fisher information $\dot{\cal I}$, the covariance $\Pi$ itself is converging. Instead, one could sample $\dot{\cal I}$ and then time-average $\Pi$ as in Algorithm 1; this would however result in an algorithm with two nested iterative loops that we do not describe here.

\section{\label{sec:special_cases}Special Cases}
Given a photon flux, raw contrast level, WFE drift statistics and corresponding sensitivity matrices for both wavefront sensor and coronagraph, one can bound the contrast achievable by wavefront sensing and control. This is done by numerically solving for residual WFE covariances ($P+Q$ or $\Pi$) as proposed in Sec.~\ref{sec:bounds} and illustrated in Sec.~\ref{sec:application_examples}. In this section, however, we first discuss some special cases in which the covariance Eqns.~(\ref{eq:bound}) and~(\ref{eq:Pi_equation}) have analytical solutions. Besides providing some theoretical insights, we re-derive results from the AO literature and show that our approach is consistent with and a generalization of previous work.

\subsection{\label{sub:special_discrete}Brownian Motion of Orthogonal Modes}

In the context of space-based coronagraphs, we explore the asymptotic
behavior of the bound derived in Sec.~\ref{sub:discrete_time} for
recursive estimation. In particular, it will be shown that the best
contrast is ``achieved'' in the limit of very short exposure time.
Additionally, we will draw a distinction between regimes in which
the image plane intensity is dominated by the initial speckle floor
(static speckles), wavefront instabilities (dynamic speckles) or Poisson-distributed incoherent sources (sensor noise, etc.).
We first make some simplifying assumptions which allow us to decouple the WFE modes. Then, treating each mode separately, we get analytical expressions for the closed-loop WFE covariances and contrasts in cases when the incoherent sources are negligible or in the limit of zero exposure time. 

The Brownian motion model is arguably the simplest non-stationary process which can describe non-smooth WFE drift that arises from structural deformations (for example, see Fig.~\ref{fig:RST_modes}(c) in Sec.~\ref{sub:RST}).
In that case, the open-loop covariance of the WFE increments between adjacent frames, $Q$, is proportional to the sampling time, $t_s$. This can be expressed as $Q=t_{s}\Xi$, where $\Xi$ is a diffusion matrix which is a property of just the wavefront instabilities.

Note that we have the freedom to choose both $Q$ and the basis of the WFE modes (the matrices $G_i$), as long as we keep the covariances of the increments for the electric fields, $G_iQG_i^T$, constant. As a result, in the monochromatic case ($c=1$) we may, without loss of
generality, choose orthognal WFE modes whose drift is uncorrelated,
\begin{equation}
Q=\begin{bmatrix}q_{1}^{2} & \cdots & 0\\
\vdots & \ddots & \vdots\\
0 & \cdots & q_{r}^{2}
\end{bmatrix},\:\Xi=\begin{bmatrix}\xi_{1}^{2} & \cdots & 0\\
\vdots & \ddots & \vdots\\
0 & \cdots & \xi_{r}^{2}
\end{bmatrix},\:\underset{i}{\sum}G_{i}^{T}G_{i}=\begin{bmatrix}\Lambda_{1}^{2} & \cdots & 0\\
\vdots & \ddots & \vdots\\
0 & \cdots & \Lambda_{r}^{2}
\end{bmatrix}\label{eq:diagonal_assumption}
\end{equation}
(since the symmetric matrix $GQG^{T}$, with $G=\begin{bmatrix}G_{1}^{T} & G_{2}^{T} & \cdots\end{bmatrix}^{T}$,
always has an orthogonal decomposition). Here $\Lambda_{j}$ can stand
for either the sensitivity at the wavefront sensor, $\Lambda_{j}^{WS}$,
or at the image plane, $\Lambda_{j}^{IP}$ .

The major assumption in this subsection is that the WFE modes are ``easily distinguishable'' by the sensor. Formally, the assumption is that the Fisher information matrix ${\cal I}$ has no cross (off-diagonal) terms. Hence, the steady-state closed-loop WFE modes are also not correlated (as a consequence of Eq.~(\ref{eq:bound}) with diagonal $Q$ and ${\cal I}$),
\[
{\cal I}=\begin{bmatrix}{\cal I}_{1} & \cdots & 0\\
\vdots & \ddots & \vdots\\
0 & \cdots & {\cal I}_{r}
\end{bmatrix},\:P=\begin{bmatrix}p_{1}^{2} & \cdots & 0\\
\vdots & \ddots & \vdots\\
0 & \cdots & p_{r}^{2}
\end{bmatrix}.
\]

To find bounds on the error variances $p_j$, we start from the approximate Eq.~(\ref{eq:EI_approx}),
and make a further simplifying assumption by replacing the summation of fractions by a fraction of summations.
This gives yet another approximation of the Fisher information,

\begin{equation}
{\cal I}_{j}\approx4\dot{N}_{S}t_{s}\frac{\underset{l=1}{\overset{r}{\sum}}(p_{l}^{2}+q_{l}^{2})\Lambda_{l}^{2}+\frac{1}{2}\left\Vert \mathbf{E}_{0}\right\Vert ^{2}}{2\underset{l=1}{\overset{r}{\sum}}(p_{l}^{2}+q_{l}^{2})\Lambda_{l}^{2}+\left\Vert \mathbf{E}_{0}\right\Vert ^{2}+\dot{N}_{S}^{-1}D}\Lambda_{j}^{2},\label{eq:diagonalized_information}
\end{equation}
where $\left\Vert \mathbf{E}_{0}\right\Vert ^{2}=\underset{i}{\sum}\left\Vert \mathbf{E}_{0,i}\right\Vert ^{2}$
and $D=\underset{i}{\sum}D_{i}$.
The WFE variances, $p_{j}^{2}$, are the solutions of Eq.~(\ref{eq:bound}),
which now take a diagonal form,

\begin{equation}
p_{j}^{-2}-(p_{j}^{2}+q_{j}^{2})^{-1}={\cal I}_{j},\label{eq:diagonalized_bound}
\end{equation}
and the contrast in Eq. (\ref{eq:contrast}) is then given by
\begin{equation}
C\approx C_{0}+2\underset{j}{\sum}\left(p_{j}^{2}+q_{j}^{2}\right)\left(\Lambda_{j}^{IP}\right)^{2}+\frac{D^{IP}}{\dot{N}_{S}^{IP}}.\label{eq:diagonalized_contrast}
\end{equation}
Equations~(\ref{eq:diagonalized_information}) and~(\ref{eq:diagonalized_bound})  for all $1\le j\le r$ are coupled (all $ p_{j}$ depend on one another),
although they become decoupled in the cases discussed below.

\subsubsection{\label{subsub:no_incoherent}Negligible incoherent sources -- recursive estimation}

One case in which we can express the contrast in terms of the system parameters is when the flux of photons from incoherent sources $D$ is negligible compared to the flux from the coherent speckles. This can be stated as $\dot{N}_{S}^{-1}D\ll\protect\underset{l=1}{\protect\overset{r}{\sum}}q_{l}^{2}\Lambda_{l}^{2}+\left\Vert \mathbf{E}_{0}\right\Vert ^{2}$. In this case, Eq.~(\ref{eq:diagonalized_information}) greatly simplifies and the information about each mode
becomes independent of the other modes, ${\cal I}_{j}\approx2\dot{N}_{S}t_{s}\Lambda_{j}^{2}$.
The estimation error variances are then given by
\[
p_{j}^{2}\approx\frac{1}{2}\left(\sqrt{1+\frac{2}{\dot{N}_{S}t_{s}\Lambda_{j}^{2}q_{j}^{2}}}-1\right)q_{j}^{2},
\]
and contrast is given by
%\begin{equation}
%C\approx C_{0}+\underset{j}{\sum}q_{j}^{2}\left(\Lambda_{j}^{IP}\right)^{2}\sqrt{1+\frac{2}{\dot{N}_{S}t_{s}\Lambda_{j}^{2}q_{j}^{2}}},
%\end{equation}
\begin{equation}
C\approx C_{0}+\underset{j}{\sum}\left(\sqrt{1+\frac{2}{\dot{N}_{S}t_{s}\Lambda_{j}^{2}q_{j}^{2}}}+1\right)\left(\Lambda_{j}^{IP}q_j\right)^{2},\label{eq:diagonalized_contrast_no_incoherent}
\end{equation}
and does not contain the negligible incoherent sources.

Since $q_{j}^{2}=\xi_{j}^{2}t_{s}$ where $\xi_j$ are some diffusion coefficients per Eq.~(\ref{eq:diagonal_assumption}), one can show that the contrast's
infimum (greatest lower bound) is at the limit $t_{s}=0$,
\begin{alignat}{1}
p_{j}^{2} & \overset{t_{s}\rightarrow0}{\longrightarrow}\frac{\xi_{j}}{\sqrt{2\dot{N}_{S}}\Lambda_{j}},\label{eq:p_random_walk}\\
C & \overset{t_{s}\rightarrow0}{\longrightarrow}C_{0}+\sqrt{\frac{2}{\dot{N}_{S}}}\underset{j}{\sum}\frac{\xi_{j}\left(\Lambda_{j}^{IP}\right)^{2}}{\Lambda_{j}}.\label{eq:C_random_walk}
\end{alignat}
This is as expected for this limiting case, as we assumed that the variance
of the measurement noise is proportional to exposure time and ignored the photon-counting confusion associated with fixed readout noise. Intuitively, if photons/electrons from all sources are Poisson distributed, one
loses information about their arrival times by increasing $t_{s}$,
thus decreasing the information rate and consequently worsening the closed loop
contrast. Since a recursive estimator remembers the measurement history, infinitely small exposures do not result in lesser information for correction updates (in total).

\subsubsection{Negligible incoherent sources -- batch estimation}

Similarly to the previous case, but with the bound in Eq.~(\ref{eq:batch_bound})
instead, the variance of the batch estimate is

\begin{equation}
\left(p_{j}^{-2}\right)_{batch}={\cal I}_{j}\approx2\dot{N}_{S}t_{s}\Lambda_{j}^{2}.\label{eq:diagonalized_p_batch}
\end{equation}
The corresponding contrast
%\[
%C_{batch}\approx C_{0}+\underset{j}{\sum}\left(\frac{1}{\dot{N}_{S}t_{s}\Lambda_{j}^{2}}+\xi_{j}^{2}t_{s}\right)\left(\Lambda_{j}^{IP}\right)^{2},
%\]
\[
C_{batch}\approx C_{0}+\underset{j}{\sum}\left(\frac{1}{\dot{N}_{S}t_{s}\Lambda_{j}^{2}}+2\xi_{j}^{2}t_{s}\right)\left(\Lambda_{j}^{IP}\right)^{2},
\]
is unbounded (becomes worse) as $t_{s}\rightarrow0$.

It is customary to optimize the sampling time for batch estimation (see
for example \cite{guyon2005limits}) by solving $dC_{batch}/dt_s=0$:
\begin{alignat*}{1}
\left(t_{s}\right)_{min} & =\sqrt{\frac{\underset{j}{\sum}\left(\frac{\Lambda_{j}^{IP}}{\Lambda_{j}}\right)^{2}}{2\dot{N}_{S}\underset{j}{\sum}\xi_{j}^{2}\left(\Lambda_{j}^{IP}\right)^{2}}},\\
\left(C_{batch}\right)_{min} & \approx C_{0}+2\sqrt{\frac{2}{\dot{N}_{S}}}\cdot\sqrt{\underset{j}{\sum}\left(\frac{\Lambda_{j}^{IP}}{\Lambda_{j}}\right)^{2}}\cdot\sqrt{\underset{j}{\sum}\xi_{j}^{2}\left(\Lambda_{j}^{IP}\right)^{2}}.
\end{alignat*}
%\begin{alignat*}{1}
%\left(t_{s}\right)_{min} & =\sqrt{\frac{\underset{j}{\sum}\left(\frac{\Lambda_{j}^{IP}}{\Lambda_{j}}\right)^{2}}{\dot{N}_{S}\underset{j}{\sum}\xi_{j}^{2}\left(\Lambda_{j}^{IP}\right)^{2}}},\\
%\left(C_{batch}\right)_{min} & \approx C_{0}+2\sqrt{\frac{1}{\dot{N}_{S}}}\cdot\sqrt{\underset{j}{\sum}\left(\frac{\Lambda_{j}^{IP}}{\Lambda_{j}}\right)^{2}}\cdot\sqrt{\underset{j}{\sum}\xi_{j}^{2}\left(\Lambda_{j}^{IP}\right)^{2}}.
%\end{alignat*}

In the hypothetical case of a single mode (e.g. $r=1$) we find an expression for optimal exposure time  that is similar to the one derived by \cite{guyon2005limits}, $\left(t_{s}\right)_{min}  = \frac{1}{ \sqrt{2}\xi_j \Lambda_{j} \dot{N}_{S}^{1/2}  }$, that captures the optimal balance between noise in sensing exposures (which decreases with $t_{s}$) and uncorrected wavefront drift during exposures (increases with $t_{s}$). Our more general analysis is thus capable of capturing the limiting cases already described in the literature. 
Note that the contrast contribution of a single mode is larger when using batch estimation when compared to recursive schemes,
\[
\left(C_{batch}\right)_{min}-C_{0}=2\left(\left(C_{recursive}\right)_{min}-C_{0}\right).
\]
%\[
%\left(C_{batch}\right)_{min}-C_{0}=\sqrt{2}\left(\left(C_{recursive}\right)_{min}-C_{0}\right).
%\]
This factor of 
%a $\sqrt{2}$
$2$
solely corresponds to the contrast improvement associated with recursive estimation for a fixed wavefront drift per WFS iteration. In practice, AO systems are limited by control lag (\cite{petit2014sphere}) neglected in this paper, which can be alleviated using predictive control. Moreover, for requirement setting exercises, such as discussed in \cite{coyle2019large}, recursive estimators enable faster sensing exposures, which turn into relaxed absolute drifts (in wavefront per unit of time).
\subsubsection{The limit $t_{s}\rightarrow0$ (recursive estimation)}

\begin{figure}
\includegraphics[scale=0.75]{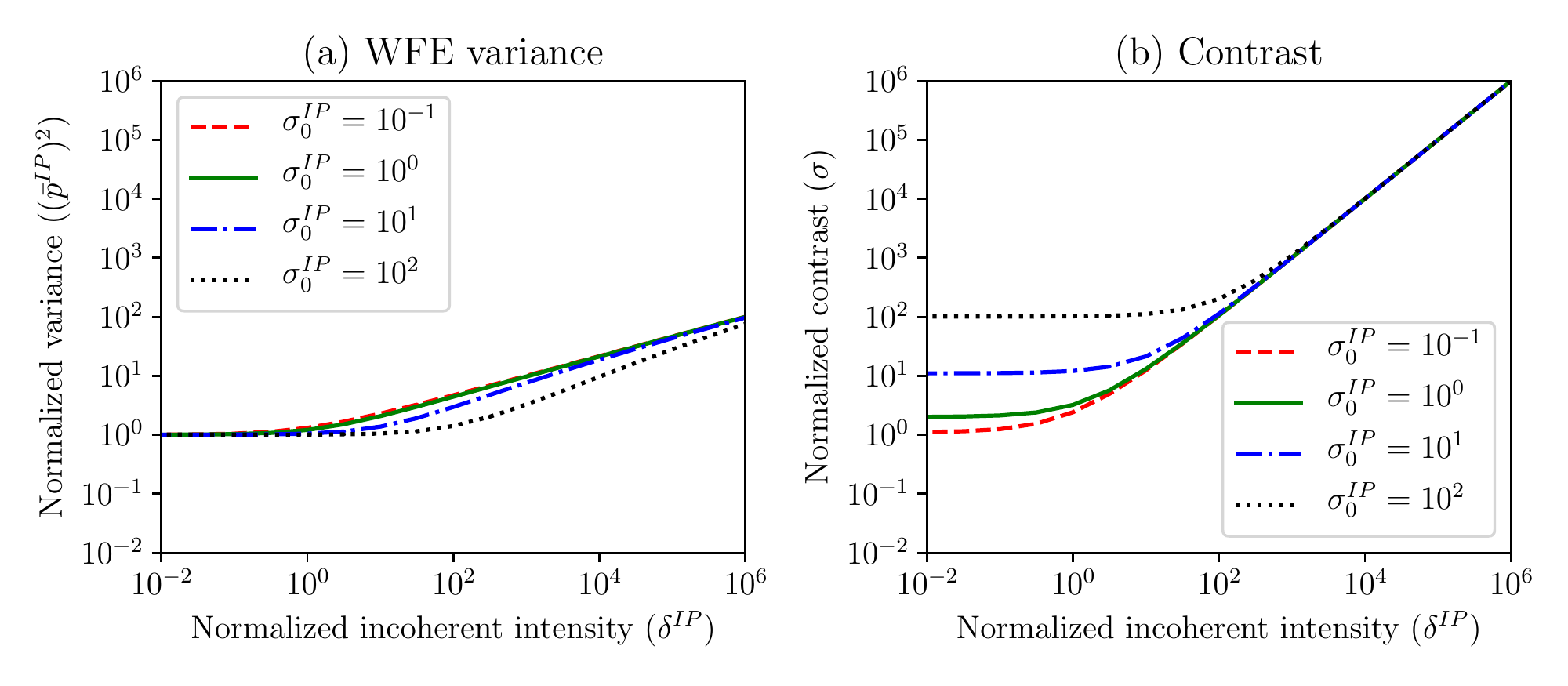}\caption{\label{fig:cubic} Image-plane wavefront control in the limit of zero exposure time, $t_{s}\rightarrow0$ (all quantities are normalized). (a) The bound on the WFE variance,
$(\bar{p}^{IP})^{2}$, grows as $\sqrt[3]{\delta^{IP}}$ when the incoherent intensity is dominant, $\delta^{IP}\gg1$, although this growth is
delayed by stronger (known) static speckles, $\sigma_{0}^{IP}$.
(b) The contrast, per Eq.~(\ref{eq:scaled_contrast}), in
three regimes: dominated by static speckles ($\delta^{IP}\ll 1$, black dotted line),
by dynamic speckles ($\delta^{IP}\ll 1$, dashed red line) or by incoherent sources
($\delta^{IP}\gg 1$).}
\end{figure}

Analytical solutions for the limiting contrast formalism can also be found in the presence of non-negligible incoherent sources ($D>0$), assuming that they are zero mean (e.g. their systematic component has been subtracted via preliminary detector calibrations) and their stochastic component follows a Poisson distribution.  We prove in Appendix~\ref{sec:optimality} that in this case, the best contrast is still achieved as $t_{s}\rightarrow0$ when using recursive estimators. Intuitively, this means that when every photon is counted individually, longer exposure times increase the probability of confusion between arrival times of distinct photons which leads to loss of information and less accurate wavefront estimation. From the hardware perspective, this $t_{s}\rightarrow0$ regime requires sensors whose readout noise decreases with exposure time. Ideally, each photon's arrival time would be tagged (see \cite{meeker_2018}).
%e.g. {\color{red}TODO for Laurent: add examples/citations for zero-exposure time sensors; also Kerri suggested expanding the discussion on clock stability}.
Whether a particular detector+estimation algorithm can operate close to the $t_{s}=0$ limit depends on its implementation and the expected number of photons per measurement. The full discussion is beyond the scope of this paper, but we suspect that propagating the full conditional probability distribution of WFE modes is stable (albeit computationally infeasible) for arbitrarily small $t_{s}$.

We will now assume $D>0$ and take the limit of Eqs.~(\ref{eq:diagonalized_information})
and (\ref{eq:diagonalized_bound}) as $t_{s}\rightarrow0$. We seek to solve for all the WFE variances $p_j$. To do this, we write the wavefront drift as $q_{j}^{2}=\xi_{j}^{2}t_{s}$, combine Eqs.~(\ref{eq:diagonalized_information}) and (\ref{eq:diagonalized_bound}), and consider the limiting case $t_{s}\rightarrow0$, giving
\begin{equation}
\xi_{j}^{2}p_{j}^{-4}=4\dot{N}_{S}\frac{\underset{l}{\sum}p_{l}^{2}\Lambda_{l}^{2}+\frac{1}{2}\left\Vert \mathbf{E}_{0}\right\Vert ^{2}}{2\underset{l}{\sum}p_{l}^{2}\Lambda_{l}^{2}+\left\Vert \mathbf{E}_{0}\right\Vert ^{2}+\dot{N}_{S}^{-1}D}\Lambda_{j}^{2}.\label{eq:diagonal_zero_t}
\end{equation}
The presence of D in the denominator of the right hand side of Eq.~(\ref{eq:diagonal_zero_t}) precludes the simplifications carried out in Sec.~\ref{subsub:no_incoherent}. However, we can still decouple this equation using the following change of variables,
\begin{equation}
\bar{p}_{j}^{2}\equiv\sqrt{2\dot{N}_{S}}\frac{\Lambda_{j}}{\xi_{j}}p_{j}^{2},\:\sigma_{0}\equiv\frac{\sqrt{\dot{N}_{s}}}{\sqrt{2}\underset{l}{\sum}\xi_{l}\Lambda_{l}}\left\Vert \mathbf{E}_{0}\right\Vert ^{2},\:\delta\equiv\frac{D}{\sqrt{2\dot{N}_{S}}\underset{l}{\sum}\xi_{l}\Lambda_{l}}.
\end{equation}
In this  modified space: all quantities are normalized by the photon rate. The closed loop variance of each individual mode is also normalized by its drift and wavefront sensor sensitivity, and both the static and incoherent intensities are normalized by the cumulative effect of all modal drifts at the wavefront sensor. Because all of these quantities are scaled by the  speckle drift, the limiting case $\sigma_{0},\delta\ll1$ corresponds to drift dominated observations (negligible incoherent noise and static contrast). After some algebra, it can be shown by direct substitution that the $r$ coupled equations given by Eq.~(\ref{eq:diagonal_zero_t}) $\forall j$, are equivalent to $r$ un-coupled equations, which we write as a single cubic equation in $\bar{p}_{j}=\bar{p},\:\forall j$,
\begin{equation}
\bar{p}^{6}+\sigma_{0}\bar{p}^{4}-\bar{p}^{2}-\sigma_{0}-\delta=0.\label{eq:p_bar_equation}
\end{equation}

The effect of the incoherent sources (including measurement noise)
now becomes apparent. When it is absent, $\delta\ll1$, $\bar{p}^{2} = 1$ is a direct solution of  Eq.~(\ref{eq:p_bar_equation}). Consequently, the estimation error, and thus closed loop variance, converges to the value in Eq.~(\ref{eq:p_random_walk}) regardless
of the magnitude of the static intensity, $\sigma_{0}$. When incoherent sources are dominant, $\delta\gg\max\{1,\sigma_{0}\}$, the variance increases proportionally to its cubic root, $\bar{p}^{2}\sim\sqrt[3]{\delta}$. These two asymptotic regimes can be identified on Fig.~\ref{fig:cubic} that illustrates the fundamental limits in normalized wavefront closed loop variance and associated contrast. Note that for Figure~\ref{fig:cubic}(b) we have assumed that the loop is closed in the image plane ($\Lambda_{j}=\Lambda_{j}^{IP}$,
$\mathbf{E}_{0}=\mathbf{E}_{0}^{IP}$), and defined the normalized contrast as
\begin{equation}
\sigma=\sigma_{0}^{IP}+(\bar{p}^{IP})^{2}+\delta^{IP}=\frac{\sqrt{\dot{N}_{s}}}{\sqrt{2}\underset{l}{\sum}\xi_{l}\Lambda_{l}^{IP}}C.\label{eq:scaled_contrast}
\end{equation}
Figure~\ref{fig:cubic}(b) shows that contrast can be limited by either
static intensity/speckles ($\sigma_{0}^{IP}\gg\max\{1,\delta^{IP}\}$, left hand side, top two curves), incoherent
sources ($\delta^{IP}\gg\max\{1,\sigma_{0}^{IP}\}$, right hand side) or wavefront instabilities
($\max\{\sigma_{0}^{IP},\delta^{IP}\}\ll1$, left hand side, bottom curve). The \emph{post-processing} contrast,
however, will be affected differently by the time varying speckles
than by the static speckles or by the constant incoherent flux. We
leave the analysis of the post-processing contrast for future work.

\subsection{\label{sub:special_AO}Higher-order Drift of a Single Mode}
\iffalse
Single complex mode ($r=2$), single channel $c=1$ and $\left\Vert G_{i}^{WS}\mathbf{\epsilon}^{CL}\right\Vert \ll\left\Vert \mathbf{E}_{0,i}^{WS}\right\Vert $
(the intensity at the wavefront sensors doesn't change much). Under
these assumptions, $\beta_{P}$ as defined in \cite{guyon2005limits} is given
by

\[
\beta_{P}=\sqrt{\mathrm{trace}\left\{ \left(\frac{\underset{i}{\sum}\frac{4}{\left\Vert \mathbf{E}_{0,i}^{WS}\right\Vert ^{2}}\left(G_{i}^{WS}\right)^{T}\mathbf{E}_{0,i}^{WS}\left(\mathbf{E}_{0,i}^{WS}\right)^{T}G_{i}^{WS}}{\underset{i}{\sum}\left\Vert \mathbf{E}_{0,i}^{WS}\right\Vert ^{2}}\right)^{-1}\right\} }
\]
(see appendix~\ref{sec:beta_p} for proof).
\fi

\subsubsection{Assumptions for continuous time}

We now consider the case of continuous time. The Brownian motion description of drifts, in which the open loop variance increases linearly as sensing exposure time, implicitly assumes a $1/f^2$ underlying power spectral density (PSD) of wavefront noise. It is a narrow assumption that, for instance, is not readily applicable to ground-based AO systems that seek to correct for atmospheric turbulence.  Here, we apply the tools described in Section \ref{sec:bounds} to derive semi-analytical contrast limits for  AO systems, or any WFS system correcting continuous time disturbances, and compare those to realistic end-to-end closed loop simulations. We derive approximate scaling laws for the dependency of the closed-loop contrast on WFE drift PSD slope and star brightness. In this section we describe the general procedure underlying these derivations, but leave out the most technical details. We  summarize our results in Table~\ref{tab:power_laws} and, similarly to \cite{males2018predictive}, we arrive at the conclusion that estimators/controllers that take into account higher-order WFE dynamics are more accurate than low-order controllers (batch estimators) and exhibit more favorable scaling laws.

For the remainder of this section we ignore realistic effects such as incoherent sources, AO-loop time delays and spatio-temporal coupling between WFE modes. To keep this exercise tractable, we consider a single real mode ($r=1$) with some open-loop PSD that decays as $f^{2\gamma}$ and is equal to $\theta^{2}$ when $f \rightarrow 0$. We wish to derive the relationship between closed loop contrast and open-loop PSD (described by $\gamma$, $\theta$, and $f_0$), the  WFE sensitivities $\Lambda^{WS}$ and $\Lambda^{IP}$, and fluxes. Note that here we distinguish between the flux at the wavefront sensor $\dot{N}_S^{WS}$ and at the image plane $\dot{N}_S^{IP}$, since they are typically not in the same band in the context of ground-based AO. The temporal PSD for this mode is written as:
\begin{equation}
\mathrm{PSD}^{OL}(f)=\left(\frac{\theta}{\left(1+\frac{f}{f_{0}}\right)^{\gamma}}\right)^{2},\:\gamma\in\mathbb{N}
\label{eq:OL_PSD}
\end{equation}
which corresponds to a $\gamma$-th order low-pass filter applied to white noise. Again, we start with the information rate in the absence of incoherent sources, and write the continuous time equivalent (i.e., $\dot{\cal I} \sim \frac{d \cal I}{d t_s} $) of Eq.~(\ref{eq:diagonalized_information}),
\[
\dot{{\cal I}}=2\dot{N}_{S}^{WS}\left(\Lambda^{WS}\right)^{2}.
\]
Note that this information rate does not depend on the static contrast, due to the peculiar property of Poisson distribution whose information doesn't depend on the magnitude of the underlying electric field (the trace of ${\cal I}$ in Eq.~(\ref{eq:information}), assuming $D_i=0$). In principle, deriving contrast limits in the continuous case can be achieved by injecting this expression for the Fischer information into Eq.~(\ref{eq:Pi_equation}) and solving for $\Pi_{1,1}$. When $r = 1$, this exercise is tractable analytically, however it becomes increasingly technical as the steepness of the PSD power law $\gamma$ increases.

\begin{table}
\begin{center}
\caption{\label{tab:power_laws}Scaling laws for closed-loop contrast contribution
($\Delta C$) of a single WFE mode with open-loop PSD described by
Eq.~(\ref{eq:OL_PSD}) (additionally, $\Delta C$ is proportional to $f_0$ in all cases). The scaling laws have been previously derived in \cite{guyon2005limits} for batch estimator with $\gamma\ge2$ and are shown in Fig.~5 of \cite{douglas2019laser} for the simple integrator with $\gamma=\frac{1}{2}\alpha=1$. We only prove the theoretical bound for integer $\gamma$.}
\begin{tabular}{|c|c|c|c|c|c|}
\hline 
 & Optimal  & \multicolumn{2}{c|}{Photon Flux ($\dot{N}_{S}$)} & \multicolumn{2}{c|}{Drift PSD ($\theta^{2}$)}\tabularnewline
\cline{3-6} 
 & Exposure & $\gamma=1$ & $\gamma\ge2$ & $\gamma=1$ & $\gamma\ge2$\tabularnewline
\hline 
Batch Estimation & $t_{s}\ne0$ & $\Delta C\propto\dot{N}_{S}^{-\frac{1}{2}}$ & $\Delta C\propto\dot{N}_{S}^{-\frac{2}{3}}$ & $\Delta C\propto\theta$ & $\Delta C\propto\theta^{\frac{2}{3}}$\tabularnewline
\hline 
Simple Integrator & $t_{s}=0$ & $\Delta C\propto\dot{N}_{S}^{-\frac{1}{2}}$ & $\Delta C\propto\dot{N}_{S}^{-\frac{2}{3}}$ & $\Delta C\propto\theta$ & $\Delta C\propto\theta^{\frac{2}{3}}$\tabularnewline
\hline 
Theoretical Bound & $t_{s}=0$ & \multicolumn{2}{c|}{$\Delta C\propto\dot{N}_{S}^{\frac{1}{2\gamma}-1}$} & \multicolumn{2}{c|}{$\Delta C\propto\theta^{\frac{1}{\gamma}}$}\tabularnewline
\hline 
\end{tabular}
\end{center}
\end{table}

\subsubsection{Continuous time Brownian motion}

%{\bf We start consider case of $\gamma=1$, that was addressed above in the discrete case and in the ``pure-integrator'' regime (defined later on in this section)}. 

For the sake of clarity, we first tackle the $\gamma = 1$ continuous case, which can be treated using a simple extension of our previous results. We follow the derivation in Section 3.1, this time using a continuous time formulation for the drift: $\xi^2 \sim \frac{d q^2}{d t_s} $. It can be shown by substituting $\xi_j$ in Eq.~(\ref{eq:C_random_walk}) with its expression as a function of $\theta$ and $f_0$ (given $\theta=f_{0}^{-1}\xi_j$), that the continuous formulation of our fundamental limit case is
\[
\Delta C_{recursive}=C-C_{0}=\left(\frac{\Lambda^{IP}}{\Lambda^{WS}}\right)^{2}\cdot\frac{f_{0}}{\dot{N}_{S}^{WS}}\cdot\left(2 \dot{N}_{S}^{WS}\theta^{2}\left(\Lambda^{WS}\right)^{2}\right)^{\frac{1}{2}},
\]
and
\[
\left(\Delta C_{batch}\right)_{min}=2\Delta C_{recursive}.
\]
%\[
%\left(\Delta C_{batch}\right)_{min}=\sqrt{2}\Delta C_{recursive}.
%\]
To simplify notation, we denote 
\begin{equation}
\bar{\theta}^{2}\equiv 2\dot{N}_{S}^{WS}\theta^{2}\left(\Lambda^{WS}\right)^{2},\:\:\Delta\bar{C}\equiv\left(\frac{\Lambda^{WS}}{\Lambda^{IP}}\right)^{2}\frac{\dot{N}_{S}^{WS}}{f_{0}}\Delta C,\label{eq:normalized_PSD}
\end{equation}
where $\bar{\theta}^2$ is the drift intensity normalized by WFE sensitivity ($\Lambda^{WS}$) and flux ($\dot{N}_S^{WS}$; note that $\dot{N}_S^{IP}$ cancels out), and $\Delta \bar{C}$ is the contrast ``contribution'' normalized by the ratio of WFE sensitivities (wavefront sensor and image plane) and by the ratio of flux to WFE PSD knee frequency ($f_0$). Just as we did in Section 3.1, we now use these normalized quantities for the remainder of this section. 

%Note that for $\gamma=1$ (Brownian motion), $\Delta \bar{C}_{recursive}=\bar{\theta}=\frac{1}{\sqrt{2}}\left(\Delta \bar{C}_{batch}\right)_{min}$.

\subsubsection{Higher order power laws -- batch estimation}

We now consider the more general case of $\gamma\ge2$ and first address theoretical bounds in the case of a batch estimator. Calculations in this case are analogous to our derivations using a discrete time formulation. That is, the contrast limit can be calculated by balancing the information content in the sensing exposure with the stochastic drift occurring during that duration,
\[
\Delta C_{batch}=2\left(p_{batch}^{2}+u(\theta, f_0,\gamma)^2 t_{s}^2\right)\left(\Lambda^{IP}\right)^{2}.
\]
%\[
%\Delta C_{batch}=2\left(p_{batch}^{2}+\frac{1}{2} u(\theta, f_0,\gamma)^2 t_{s}^2\right)\left(\Lambda^{IP}\right)^{2}.
%\]
Now that open loop variance is not an affine function of time, we consider the average stochastic drift $ u(\theta, f_0,\gamma) = < |d \epsilon / dt| >$, which is the only relevant quantity when using a batch estimator that averages out higher order wavefront dynamics. Using dimensional analysis, one can show that this average drift scales as
\[
u(\theta, f_0,\gamma)^2  = a_{\gamma}^2 \theta^{2}f_{0}^{3},
\]
where $a_{\gamma}$ is some dimensionless constant. The contrast contribution of the single mode with batch estimation is thus
\[
\Delta C_{batch}=2\left(\frac{1}{\dot{N}_{S}^{WS}t_{s}\left(\Lambda^{WS}\right)^{2}}+ a_{\gamma}^2\theta^2 f_{0}^{3} t_{s}^2\right)\left(\Lambda^{IP}\right)^{2},
\]
%\[
%\Delta C_{batch}=2\left(\frac{1}{\dot{N}_{S}^{WS}t_{s}\left(\Lambda^{WS}\right)^{2}}+ \frac{1}{2}a_{\gamma}^2\theta^2 f_{0}^{3} t_{s}^2\right)\left(\Lambda^{IP}\right)^{2},
%\]
where we used Eq.~(\ref{eq:diagonalized_p_batch}) to substitute the closed loop variance, $p_{batch}^2$, with the Fischer information. By differentiating w.r.t. $t_s$, the minimum contribution is, up to some constant (see also \cite{guyon2005limits}),

%\begin{equation}
%\left(\Delta C_{batch}\right)_{min}\sim\left(\frac{\Lambda^{IP}}{\Lambda^{WS}}\right)^{2}\cdot\f%rac{f_{0}}{\dot{N}_{S}^{WS}}\cdot\left(\dot{N}_{S}^{WS}\theta^{2}\left(\Lambda^{WS}\right)^{2}\r%ight)^{\frac{1}{3}}\label{eq:batch_power_law}
%\end{equation}
\[
\left(\Delta \bar{C}_{batch}\right)_{min} \sim \bar{\theta}^{\frac{2}{3}}
\]
and is obtained when
\[
\left(t_{s}\right)_{min} \sim \bar{\theta}^{-\frac{2}{3}} f_{0}^{-1}.
\]

\begin{figure}
\includegraphics[scale=0.75]{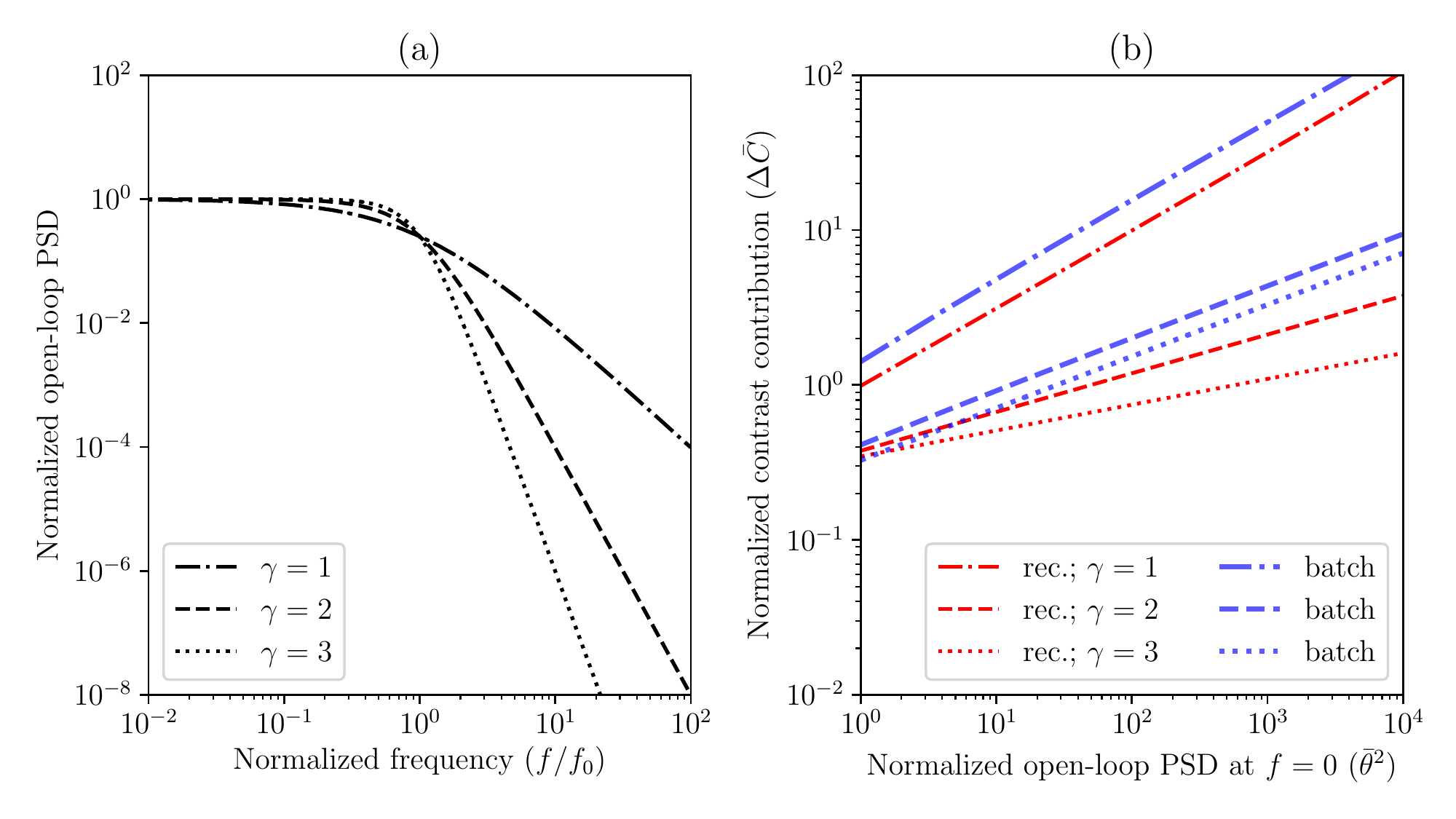}\caption{\label{fig:power_laws}(a) Open-loop WFE PSD corresponding to Eq.~(\ref{eq:OL_PSD})
and various values of $\gamma$. For $\gamma=1$ it is the Ornstein-Uhlenbeck
process (\cite{durrett2019probability}). (b) The dependency of normalized closed-loop contrast on
drift magnitude in the $\bar{\theta}^{2}=\theta^{2}\dot{{\cal I}}\gg1$
limit. For $\gamma=1$, recursive estimation is more accurate than
batch estimation but their scaling is the same. For $\gamma\ge2$
(when the WFE mode is differentiable in time), recursive estimation
is both more accurate and scales better with drift magnitude and photon
flux.}
\end{figure}

\subsubsection{Higher order power laws -- recursive estimation}

To analyze recursive estimation we cannot simply consider average drifts; we have to actually solve for $\Pi_{1,1}$ in Eq.~(\ref{eq:Pi_equation}). To do so, we note that the PSD in Eq.~(\ref{eq:OL_PSD}) corresponds
to an integrator of order $\gamma$ of white noise, $v(t)$ s.t. $\underset{t}{\overset{t+\Delta t}{\int}}vdt\sim{\cal N}\left(0,\Delta t\right)$. Stated in terms of Eq. (\ref{eq:continous}) the PSD corresponds to,

\begin{equation}
\begin{bmatrix}\frac{d}{dt}\epsilon\\
\frac{d^{2}}{dt^{2}}\epsilon\\
\vdots\\
\frac{d^{\gamma}}{dt^{\gamma}}\epsilon
\end{bmatrix}=\underset{A(f_0)}{\underbrace{\begin{bmatrix} -f_0 & 1\\
 & \ddots & \ddots\\
 &  & \ddots & 1\\
 &  &  & -f_0
\end{bmatrix}}}\begin{bmatrix}\epsilon\\
\frac{d}{dt}\epsilon\\
\vdots\\
\frac{d^{\gamma-1}}{dt^{\gamma-1}}\epsilon
\end{bmatrix}+\underset{B}{\underbrace{\begin{bmatrix}0\\
\vdots\\
0\\
\theta f_{0}^{\gamma}
\end{bmatrix}}}v(t).\label{eq:drift_state_equation}
\end{equation}

We consider the regime for which $ f \gg f_0$. This corresponds to the case of  short wavefront sensing  timescales (much shorter than $f^{-1}_{0}$), that will benefit most from using a recursive estimator; for longer timescales using a batch estimator is sufficient. For short timescales, the solutions of Eq.~(\ref{eq:Pi_equation}) are insensitive to the precise values of $f_0$ in the $A$ matrix and WFE covariance (first block of the full state covariance $\Pi$) is given, up to some constant, by
\[
\Pi_{1,1}\sim\left(\theta^{2}f_{0}^{2\gamma}\dot{{\cal I}}^{1-2\gamma}\right)^{\frac{1}{2\gamma}}.
\]
For brevity, we leave out of this paper the somewhat technical proof of this relation for integer $\gamma$ (the proof does not hold for non-integer $\gamma$, but we use this relation for comparison purposes in the remainder of this section nevertheless). Plugging this expression into the continuous time expression for the contrast limit without incoherent noise, $\Delta C_{recursive}=2\Pi_{1,1}\left(\Lambda^{IP}\right)^{2}$, we find the normalized scaling law
\begin{equation}
\Delta \bar{C}_{recursive}\sim \bar{\theta}^{\frac{1}{\gamma}}\label{eq:recursive_power_law},
\end{equation}
and therefore the (not normalized) contrast satisfies the following proportionalities
\[
\Delta C_{recursive} \propto \dot{N}_{S}^{\frac{1}{2\gamma}-1},\:\Delta C_{recursive} \propto f_0.
\]

\subsubsection{Summary of analytical results in the continuous case}

The scaling laws derived in the continuous case are summarized in Table~\ref{tab:power_laws} and illustrated in Figure~\ref{fig:power_laws}. A few broad conclusions can be drawn from this work.  In all cases, recursive estimation with (close to) zero exposure time is more accurate than batch estimation with its corresponding optimal exposure time. Naturally, the normalized closed-loop contrast, $\Delta \bar{C}$, increases with the normalized drift magnitude $\bar{\theta}^2$ for various open-loop WFE PSDs of the form of Eq.~(\ref{eq:OL_PSD}). While batch estimation becomes more accurate if WFE drift is once differentiable ($\gamma=2$), it exhibits the same scaling for smoother dynamics ($\gamma\ge2$). However, the bound on recursive estimation always scales more favorably with higher orders of the drift dynamics such as the $\gamma=\frac{17}{6}$ associated with von K\'arm\'an turbulence (\cite{hardy1998adaptive}).

\subsubsection{Comparison with end-to-end Adaptive Optics simulations }

\begin{figure}

\includegraphics[scale=0.75]{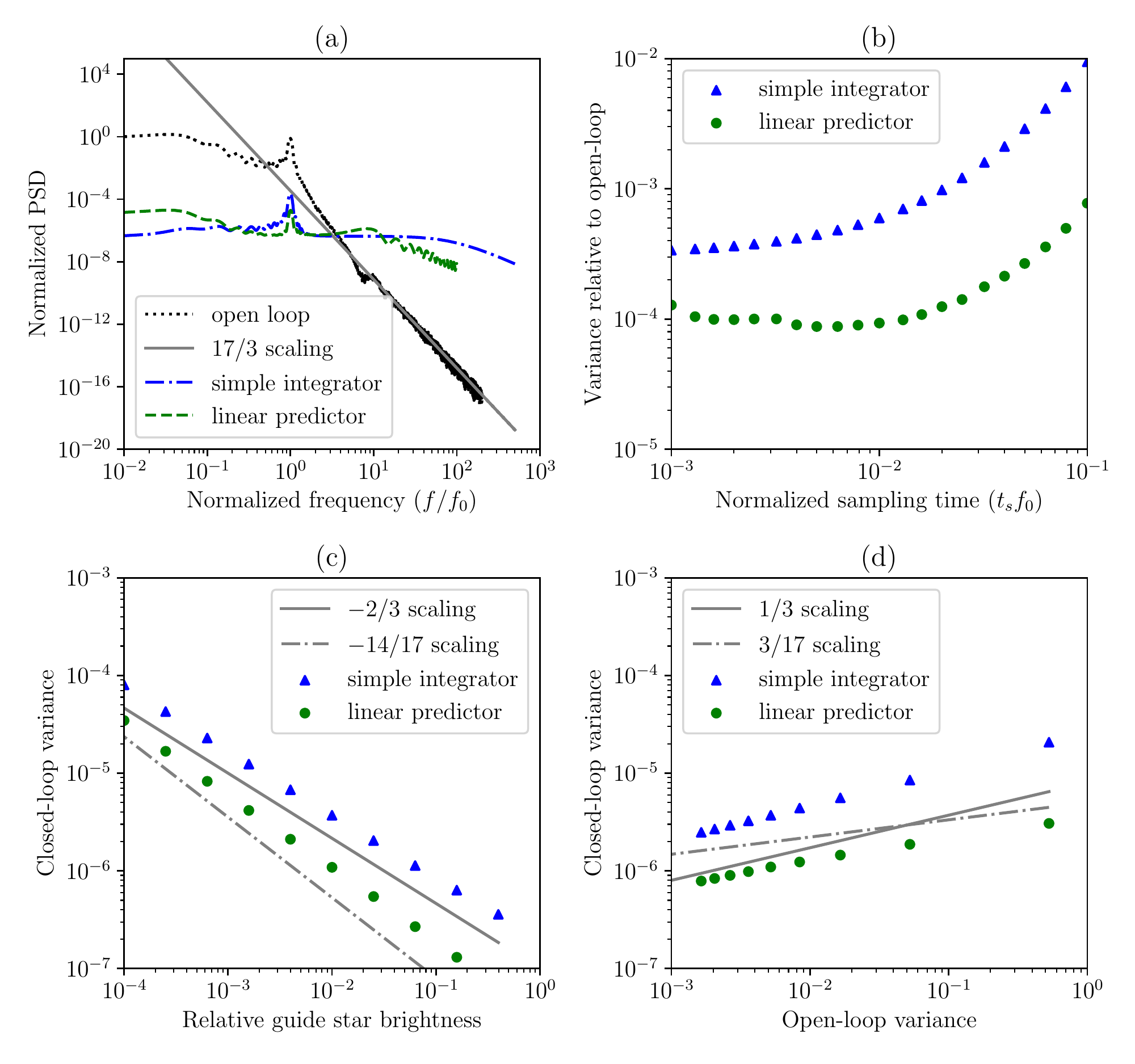}\caption{\label{fig:ao}(a) Numerically simulated PSD of a single WFE mode before (black dotted line)
and after AO correction (dash-dotted blue line for simple integrator (SI)
estimator and dashed green line for ``linear predictor'' (LP) (\cite{males2018predictive})).
The open-loop PSD scales as $f^{-\frac{17}{3}}$ ($\gamma=\frac{17}{6}$)
in the limit $f\rightarrow\infty$. (b) The closed-loop variance of the SI decreases with exposure time since it is a recursive estimator. On the other hand, the higher-order LP is more accurate but exhibits a non-zero optimal exposure time which suggests that it is not ``purely'' recursive. (c) and (d) The scaling of the closed-loop contrast of the SI (blue triangles) matches the analytically derived Expression~(\ref{eq:SI_scaling}) slopes represented by solid gray lines with arbitrary offsets). The LP (green circles) scales more favorably than the SI but not as well as the theoretical bound for recursive estimation (slopes represented by dash-dotted gray lines).}

\end{figure}
 We now compare the scaling laws resulting from our derivations (summarized in Table~\ref{tab:power_laws}) to more realistic Adaptive Optics simulations. So far we have  worked under the assumption that the sole source of closed loop variance is noise in the wavefront estimate. Comparisons with end-to-end simulations require also including errors stemming from the control law.
%As illustrated in Fig.~\ref{fig:power_laws}, batch estimators suffer from poor contrast scaling laws because they discard previous measurements. 
In this context we first discuss the performance of the simple integrator (SI) -- the simplest control
law that incorporates all measurements. We then compare SI to a linear predictor (LP) (\cite{males2018predictive}) and the bound in Expression~(\ref{eq:recursive_power_law}).

With some abuse of notation, we define the SI AO control/estimation law as
\[
\hat{\epsilon}(t)\propto\underset{0}{\overset{t}{\int}}\Delta\dot{y}(\tau)d\tau
\]
where $\hat{\epsilon}$ is the estimate of the WFE mode and $\Delta \dot{y}$ is the measured deviation of the intensity from its nominal flat-wavefront value at the WFS (even though this would classically be considered a control rather than estimation law, there is effectively no distinction between the two since we assumed direct influence of the DM on the WFE with no time varying irregularities).
In terms of the normalized quantities in Eq.~(\ref{eq:normalized_PSD}), we show in Appendix~\ref{sub:SI_PSD} that the closed-loop contrast contribution with the SI is, up to some constant,
\begin{equation}
\Delta\bar{C}^{CL,SI}\sim\begin{cases}
\bar{\theta} & \gamma=1\\
\bar{\theta}^{\frac{2}{3}} & \gamma\ge2
\end{cases}\label{eq:SI_scaling}
\end{equation}
Note that although the SI contrast power law is the same as for batch estimation, the SI benefits from reducing the exposure time, while for batch estimators the optimal exposure time is finite. We also report these results in  Table~\ref{tab:power_laws}.

We can then compare our results to performances of the ground-based AO-fed coronagraph analyzed using the semi-analytic framework from \cite{males2018predictive}.  We filtered the temporal power spectra of Fourier modes in von K\'arm\'an turbulence (\cite{hardy1998adaptive})  by optimized control laws (see Fig.~\ref{fig:ao}(a)), and determined the post-coronagraph contrast from the residual variance in each mode assuming an ideal coronagraph. The control laws (SI and LP) were optimized to minimize variance per Fourier mode.  We varied WFS exposure times, guide star brightness, and the Fried parameter, which changes the open-loop variance.  To simplify analysis, zero loop delay was assumed, except for the sample-and-hold from finite integration.  

Figure~\ref{fig:ao}(b) shows the dependency of the residual WFE
covariance of both controllers on the sampling time. Since the simple
integrator is a recursive estimator, its accuracy becomes better with
decreasing sampling time. When $\gamma=1$, the SI is the optimal recursive estimator.
This is not the case when $\gamma >1$; for instance its first order dynamics make it
sub-optimal in the $\gamma=\frac{17}{6}$ case and it is therefore
less accurate than the higher-order linear predictor. On the other hand the non-zero
optimal sampling time of the linear predictor suggests that it is
not ``purely'' recursive. Figure~\ref{fig:ao}(c) and (d) show
that the scaling laws we derived for closed-loop variance as a function of guide star brightness and open loop variance  of the simple integrator broadly match the more sophisticated simulations in \cite{males2018predictive}. As a matter of fact,  scaling of the closed-loop contrast of the SI matches the analytically derived Expression~(\ref{eq:SI_scaling}). The LP (green circles) scales more favorably than the SI but, as expected, not as well as the theoretical bound for recursive estimation (dash-dotted gray line). This good match occurs in spite of the drastically stringent assumption underlying our analytical work. This demonstrates the potential of carrying out more detailed analyses of Adaptive Optics systems by numerically solving Eq.~(\ref{eq:Pi_equation}) and using algorithms akin to the ones presented in Section~\ref{sub:implementation}. We however leave out the execution of such investigations for future work.

\section{\label{sec:application_examples}Space-based Coronagraph Applications}

We now focus on space-based applications of our novel formulation and relate the theory in Sec.~\ref{sec:bounds} to commonly-used single-pixel based estimation in the focal plane (\ref{sub:example_single}), combining estimated bounds from LOWFS and HOWFS (\ref{sub:combining_bounds}), and estimating the closed-loop contrast of the Roman Space Telescope (\ref{sub:RST}). Consistent with the results in Sec.~\ref{sec:special_cases}, our numerical simulations show that batch estimation is less ``efficient'' than recursive estimation, that the closed-loop contrast of the RST is dominated by the incoherent sources and that its ``dynamic'' portion scales proportionally to the cubic root of the sources' combined intensity.

\subsection{\label{sub:example_single}Brownian Motion of the Electric Field
of a Single Image-plane Pixel}

In image-plane wavefront sensing and control, it is common, for estimation
purposes, to treat each detector pixel separately (\cite{give2011pair,riggs2014optimal}).
Such approaches require ``probing'' or ``dithering'' the DM to
introduce sufficient phase diversity to distinguish between real and
imaginary parts of the electric field based on intensity measurements
alone. Here we adjust the analytical bound proposed in Sec.~\ref{sub:discrete_time}
to this particular case and compare it to numerical simulations. For such applications it is common to estimate the electric field at a single pixel, $\mathbf{E}=G\bm{\epsilon}$, in coronagraph images, instead of the WFE mode coefficients, $\bm{\epsilon}$. Instead of ${\cal I}$, we can use the matrix ${\cal E}$,  which stands for the Fisher information about the electric field contained in a single photon-counting measurement and features two degrees of freedom (real and imaginary part) for each pixel.

The dynamic (closed-loop and zero-mean) component of the electric
field at the single pixel will be denoted by $\mathbf{E}^{CL}\in\mathbb{R}^{2}$.
Instead of the static part, $\mathbf{E}_{0}$, we introduce four ``probes'', $\mathbf{E}_{p}\in\left\{ \begin{bmatrix}\sqrt{C_{p}}\\
0
\end{bmatrix},\begin{bmatrix}-\sqrt{C_{p}}\\
0
\end{bmatrix},\begin{bmatrix}0\\
\sqrt{C_{p}}
\end{bmatrix},\begin{bmatrix}\\
-\sqrt{C_{p}}
\end{bmatrix}\right\} $, with equal probability (the $\cdot^{IP}$ superscript will be dropped
throughout this example). Here, $\sqrt{C_{p}}$ denotes the magnitude
of the probes (such that if $\mathbf{E}^{CL}\equiv0$, the contrast
is $C=C_{p}$).

Neglecting incoherent sources ($D=0$), the information about $\mathbf{E}^{CL}$
contained in each photon count is (similar to Eq.~(\ref{eq:information}), but dimensionless),

\[
{\cal E}\equiv4\dot{N}_{S}t_{s}\left(\frac{\mathbf{E}^{CL}+\mathbf{E}_{p}}{\left\Vert \mathbf{E}^{CL}+\mathbf{E}_{p}\right\Vert }\right)\left(\frac{\mathbf{E}^{CL}+\mathbf{E}_{p}}{\left\Vert \mathbf{E}^{CL}+\mathbf{E}_{p}\right\Vert }\right)^{T}.
\]
The random unit vector $\left(\mathbf{E}^{CL}+\mathbf{E}_{p}\right)/\left\Vert \mathbf{E}^{CL}+\mathbf{E}_{p}\right\Vert \in\mathbb{R}^{2}$
has a probability density function that is symmetric w.r.t. rotations
by $\pi/2$ , hence
\[
\mathrm{E}_{\mathbf{E},\mathbf{E}_{p}}\left\{ {\cal E}\right\} =4\dot{N}_{S}t_{s}\mathrm{cov}\left\{ \frac{\mathbf{E}^{CL}+\mathbf{E}_{p}}{\left\Vert \mathbf{E}^{CL}+\mathbf{E}_{p}\right\Vert }\right\} =2\dot{N}_{S}t_{s}\begin{bmatrix}1 & 0\\
0 & 1
\end{bmatrix}.
\]
In accordance with Sec.~\ref{sub:discrete_time}, we further split
the closed-loop electric field into drift and estimation errors,
\[
%\mathbf{E}^{CL}=\mathbf{E}(t+t_{s})-\hat{\mathbf{E}}(t)=\mathbf{E}(t+t_{s})-\mathbf{E}(t)\:\:+\:\:\mathbf{E}(t)-\hat{\mathbf{E}}(t),
\mathbf{E}^{CL}=\mathbf{E}(t+t_{s})-\hat{\mathbf{E}}(t)=\underset{\text{error due to drift}}{\underbrace{\mathbf{E}(t+t_{s})-\mathbf{E}(t)}}\:\:+\:\:\underset{\text{estimation error}}{\underbrace{\mathbf{E}(t)-\hat{\mathbf{E}}(t)}},
\]

that are assumed to be normally distributed with
\begin{alignat*}{1}
\mathrm{cov}\left\{ \mathbf{E}(t+t_{s})-\mathbf{E}(t)\right\}  & =\Lambda^{2}\xi^{2}t_{s}\begin{bmatrix}1 & 0\\
0 & 1
\end{bmatrix}=Q,\\
\mathrm{cov}\left\{ \mathbf{E}(t)-\hat{\mathbf{E}}(t)\right\}  & =p^{2}\begin{bmatrix}1 & 0\\
0 & 1
\end{bmatrix}=P.
\end{alignat*}

The bounds on batch and recursive estimation error variances are given
in Eqs.~(\ref{eq:bound}) and (\ref{eq:batch_bound}),

\begin{alignat*}{1}
p_{batch}^{2} & =\frac{1}{2\dot{N}_{S}t_{s}},\\
p_{recursive}^{2} & =\frac{1}{2}\left(\sqrt{1+\frac{2}{\dot{N}_{S}\Lambda^{2}\xi^{2}t_{s}^{2}}}-1\right)\Lambda^{2}\xi^{2}t_{s},
\end{alignat*}
and the corresponding contrasts (Eq.~(\ref{eq:contrast})) are 
\begin{alignat*}{1}
C_{batch} & =\frac{1}{\dot{N}_{S}t_{s}}+2\Lambda^{2}\xi^{2}t_{s}+C_{p},\\
C_{recursive} & =\left(\sqrt{1+\frac{2}{\dot{N}_{S}\Lambda^{2}\xi^{2}t_{s}^{2}}}+1\right)\Lambda^{2}\xi^{2}t_{s}+C_{p}.
\end{alignat*}
%\begin{alignat*}{1}
%C_{batch} & =\frac{1}{\dot{N}_{S}t_{s}}+\Lambda^{2}\xi^{2}t_{s}+C_{0},\\
%C_{recursive} & =\Lambda^{2}\xi^{2}t_{s}\sqrt{1+\frac{2}{\dot{N}_{S}\Lambda^{2}\xi^{2}t_{s}^{2}}}+C_{0}.
%\end{alignat*}
The lower bounds that are obtained when infimizing with respect to
$t_{s}$, are
\begin{alignat}{1}
\inf\left\{ C_{batch}-C_{p}\right\}  & =2\sqrt{\frac{2\Lambda^{2}\xi^{2}}{\dot{N}_{S}}}\:\mathrm{at}\:t_{s}=\sqrt{\frac{1}{2\Lambda^{2}\xi^{2}\dot{N}_{S}}},\label{eq:single_time}\\
\inf\left\{ C_{recursive}-C_{p}\right\}  & =\sqrt{\frac{2\Lambda^{2}\xi^{2}}{\dot{N}_{S}}}\:\mathrm{at}\:t_{s}=0.\label{eq:single_scaling}
\end{alignat}
%\begin{alignat}{1}
%\inf\left\{ C_{batch}-C_{0}\right\}  & =2\sqrt{\frac{\Lambda^{2}\xi^{2}}{\dot{N}_{S}}}\:\mathrm{at}\:t_{s}=\sqrt{\frac{1}{\Lambda^{2}\xi^{2}\dot{N}_{S}}},\label{eq:single_time}\\
%\inf\left\{ C_{recursive}-C_{0}\right\}  & =\sqrt{\frac{2\Lambda^{2}\xi^{2}}{\dot{N}_{S}}}\:\mathrm{at}\:t_{s}=0.\label{eq:single_scaling}
%\end{alignat}

\begin{figure}
\includegraphics[scale=0.75]{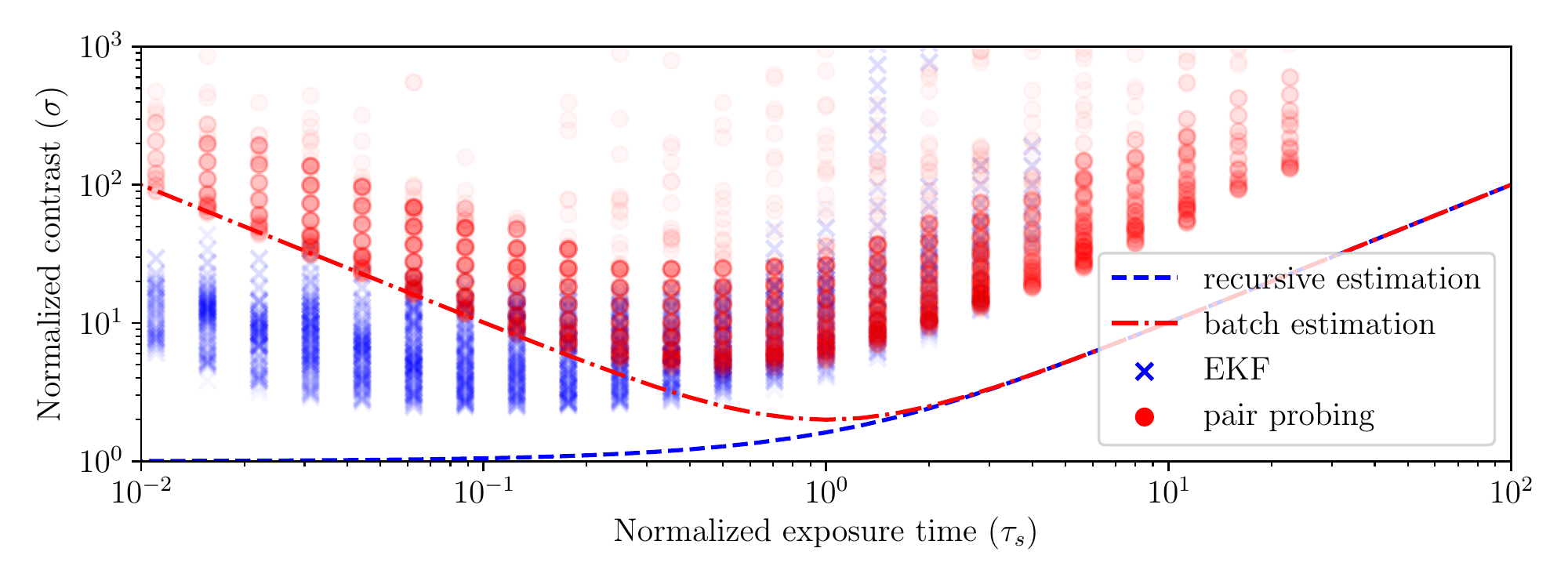}\caption{\label{fig:example_single}The closed-loop contrast at a single pixel
(normalized by the lower bound in Eq.~(\ref{eq:single_scaling})) as
a function of the sampling time (normalized by the optimal sampling time
for batch estimation in Eq.~(\ref{eq:single_time})). The analytical
bound for recursive estimation (dashed blue line) converges to its
lower limit as $\tau_{s}\rightarrow0$, while the minimum batch estimation
contrast (dash-dotted red line) is worse by a factor of $2$. The best
contrasts achieved in numerical simulations of EKF (${\color{blue}\times}$)
and pair probing (${\color{red}\bullet}$), are worse than their respective
lower bounds, but the EKF outperforms pair probing by a factor of about $2$. Eventually, as the sampling
time approaches zero, the EKF breaks down depending on the value of the
photon flux (since the Gaussian approximation of shot noise in \cite{pogorelyuk2019dark}
becomes progressively less accurate as the flux decreases).}
\end{figure}

Yet, the electric field at a single pixel cannot be estimated with
$C_{p}=0$ (no probing) due to phase ambiguity. Therefore, in order to compare our analytical limits to practical implementations, we conducted a series of simulations of pair-probing (\cite{give2011pair}) and an Extended Kalman Filter (\cite{pogorelyuk2019dark})
(EKF). In these simulations we fixed $\Lambda^{-2} \xi^{-2} = 1\:\mathrm{s}$, varied $\dot{N}_S^{-1}$ and $t_s$ between $2^{-5}\:\mathrm{s}$ and $2^5\:\mathrm{s}$, and varied $\sqrt{C_{p}}$ between $2^{-5}$ and $2^{5}$. The temporal discretization of the simulation was $2^{-8}\:\mathrm{s}$ and the photon counts during a single exposure were stacked.

Figure \ref{fig:example_single} shows the contrast ($\sigma$), normalized
by the recursive-estimation bound, as a function of sampling time ($\tau_{s}$),
normalized by the optimal sampling time for batch estimation,
\begin{alignat*}{1}
\sigma & \equiv\sqrt{\frac{\dot{N}_{S}}{2\Lambda^{2}\xi^{2}}}C,\\
%\tau_{s} & \equiv\sqrt{\Lambda^{2}\xi^{2}\dot{N}_{S}}t_{s}.
\tau_{s} & \equiv\sqrt{2\Lambda^{2}\xi^{2}\dot{N}_{S}}t_{s}.
\end{alignat*}
When the EKF was stable at short exposure times, it outperformed batch estimation by a factor
of about two. The general behavior of normalized contrast as a function of normalized exposure time does follow the expected theoretical bound for both pair-probing and EKF, up to a multiplicative constant. Indeed,  both achieved contrast worse than their corresponding
analytical bounds.  This discrepancy highlights the improvement that might be achieved when using more optimal dark hole maintenance algorithms. Moving forward, we encourage future innovations in this field to be bench-marked against our theoretical bound.

\subsection{\label{sub:combining_bounds}Combining Bounds for Low and High Order
Wavefront Sensing (Space Coronagraphs)}

So far we have considered the case of a single WFS loop correcting corrugation due to atmospheric turbulence or internal drifts. In practice, future space based observatories might operate using several nested closed loops operating at time scales spanning a few orders of magnitude. For instance, the  LOWFS loop of Roman Space Telescope will operate at $>10\:\mathrm{Hz}$ to counteract
the fast line-of-sight disturbance by the reaction wheels (\cite{shi2017dynamic}),
while higher order wavefront disturbances due to thermal deformation
of the OTA are slower by at least two orders of magnitude (\cite{krist2018wfirst}).
As a result, one cannot assume that the post-LOWFS residual WFE modes
(jitter) are quasi-static during the minutes-long exposures. Fortunately, the large separation between time scales
allows treating the jitter residual as an additional source of incoherent
light, while higher order modes remain decoupled and evolve slowly (see \cite{pogorelyuk2020effects}, and note that the incoherent intensity associated with jitter changes over time as reaction wheels build up momentum).We can thus also apply our methodology to derive theoretical bounds under this more realistic scenario. Below we outline the key steps to be undertaken to do so, but we leave comparisons between bounds and realistic simulations to a future publication.

\begin{figure}

\includegraphics{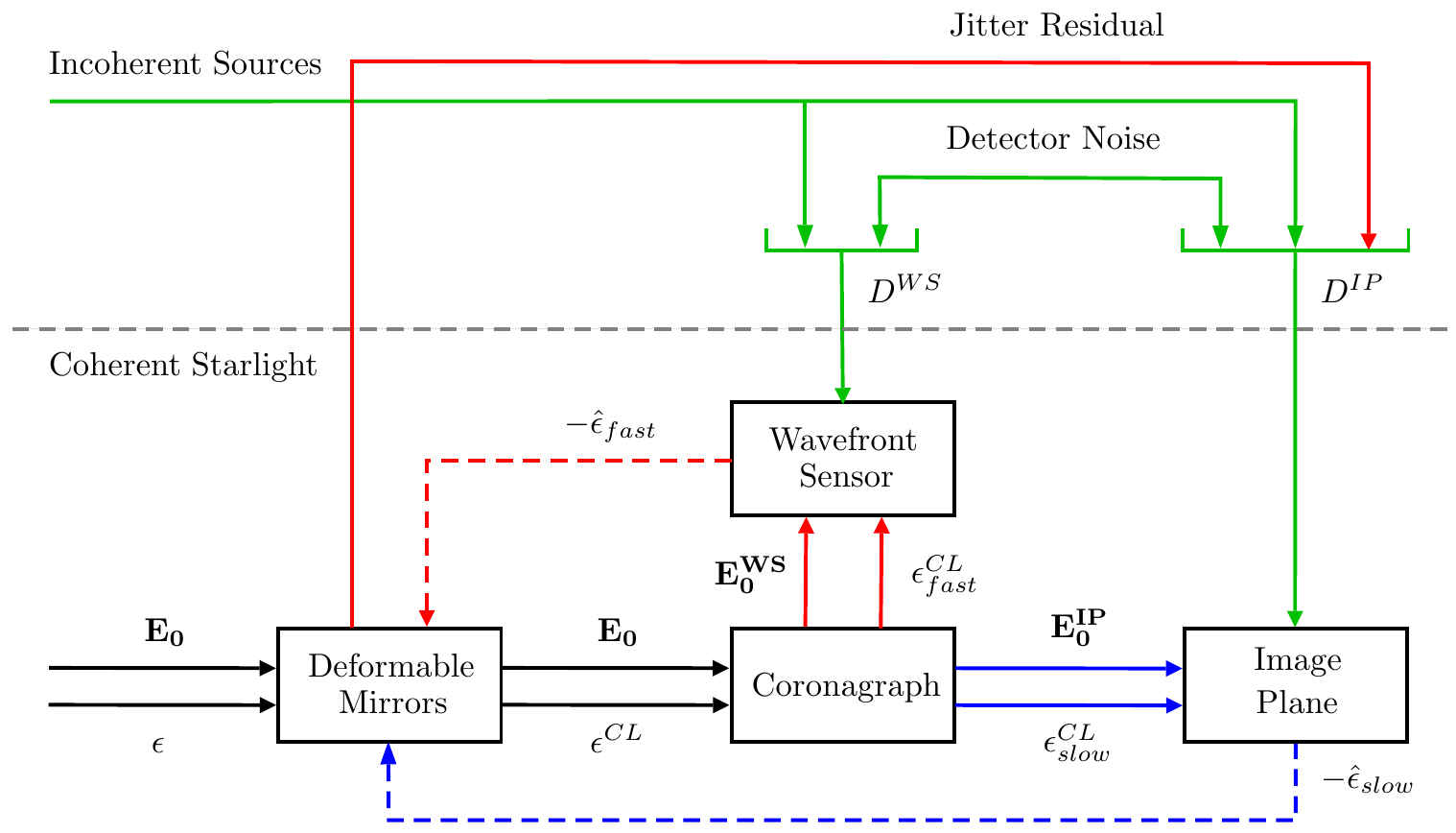}\caption{\label{fig:diagram}LOWFS and HOWFS loops of a space based coronagraph (based on the architecture in \cite{pueyo2019luvoir}), with coherent starlight (bottom) and other light sources (top) propagating from left to right.
In this simplified model, WFE modes can be conceptually split into
``fast'', handled by LOWFS in a non-common path wavefront sensor (center),
and slow, observable in the image plane (bottom left). The fast LOWFS residuals
appear as an incoherent light source in the image plane (\cite{pogorelyuk2020effects}), along with
detector noise and external sources (e.g. zodi).}
\end{figure}

We illustrate how the contrast bounds can be found sequentially,
first for LOWFS and then for HOWFS, taking into account the influence
of the former on the latter (this can be extended to  segmented telescopes that might require three nested WFS loops, some out-of-band). To simplify notations, we assume that
the WFE can be split (Fig.~\ref{fig:diagram}) into fast and slow
modes handled by LOWFS and HOWFS respectively,
\[
\bm{\epsilon}=\begin{bmatrix}\bm{\epsilon}_{fast}\\
\bm{\epsilon}_{slow}
\end{bmatrix}.
\]
This distinction is suitable, for example, for RST where the residual fast modes, $\bm{\epsilon}_{fast}^{CL}$,
have a zero mean over the sampling time of the slow loop ($t_{s,slow}$),

\[
\left\langle \bm{\epsilon}_{fast}^{CL}\right\rangle =\frac{1}{t_{s,slow}}\underset{t_{k}}{\overset{t_{k}+t_{s,slow}}{\int}}\bm{\epsilon}_{fast}^{CL}dt\approx\mathbf{0},
\]
where $k$ denotes the exposure number (otherwise, the analysis remains
valid but the notations become cumbersome). We also assume
that LOWFS and HOWFS operate in the same spectral band, hence $\dot{N}_{S}^{IP}=\dot{N}_{S}^{WS}=\dot{N}_{S}$.
We proceed by splitting the sensitivities of the wavefront sensor
at the image plane based on LOWFS (fast) and HOWFS (slow) modes

\begin{alignat*}{1}
G_{i}^{WS} & =\begin{bmatrix}G_{i,fast}^{WS} & G_{i,slow}^{WS}\end{bmatrix},\\
G_{i}^{IP} & =\begin{bmatrix}G_{i,fast}^{IP} & G_{i,slow}^{IP}\end{bmatrix},
\end{alignat*}
and ``closing the loops'' separately,

\[
\bm{\epsilon}^{CL}(t)=\begin{bmatrix}\bm{\epsilon}_{fast}^{CL}\\
\bm{\epsilon}_{slow}^{CL}
\end{bmatrix}=\bm{\epsilon}-\begin{bmatrix}\hat{\bm{\epsilon}}_{fast}^{WS}(t)\\
\left(\hat{\bm{\epsilon}}_{slow}^{IP}\right)_{k}
\end{bmatrix},\:t\in\left[t_{k},t_{k+1}\right).
\]
Here, to illustrate that different temporal analyses can be combined, $\hat{\bm{\epsilon}}_{fast}^{WS}(t)$ is treated in continuous time (similarly to AO in Sec.~\ref{sub:special_AO}),
and $\left(\hat{\bm{\epsilon}}_{slow}^{IP}\right)_{k}$ in discrete time.

The fast loop can be analyzed with the formalism presented in Sections
\ref{sub:discrete_time} or \ref{sub:continuous_time}, with the intensity
and photon counts at the wavefront sensor given by

\begin{alignat*}{1}
I_{i}^{WS} & =\dot{N}_{S}\left\Vert G_{i,fast}^{WS}\bm{\epsilon}_{fast}^{CL}+\mathbf{E}_{0,i}^{WS}\right\Vert ^{2},\\
y_{i}^{WS} & \sim poisson\left(\left(I_{i}^{WS}+D_{ext,i}^{WS}+D_{int,i}^{WS}\right)t_{s,fast}\right),
\end{alignat*}
where $D_{ext,i}^{WS}$ includes zodi, etc.~and $D_{int,i}^{WS}$ induces
dark current, etc. The covariance of the closed loop fast WFE residuals,
$\Pi_{fast}$ s.t. $\bm{\epsilon}_{fast}^{CL}\sim{\cal N}\left(\mathbf{0},\Pi_{fast}\right)$,
can be found as prescribed in Sec.~\ref{sub:implementation}.

In the image plane, the average fast WFE modes are effectively zero, $\left\langle \bm{\epsilon}_{fast}^{CL}\right\rangle =\mathbf{0}$,
hence they do not contribute to the intensity of the \emph{coherent}
speckles. However, their average intensity contribution is positive,
\[
D_{jit,i}^{IP}\equiv\left\langle \left\Vert G_{i,fast}^{IP}\bm{\epsilon}_{fast}^{CL}\right\Vert ^{2}\right\rangle =\mathrm{trace}\left\{ G_{i,fast}^{IP}\Pi_{fast}\left(G_{i,fast}^{IP}\right)^{T}\right\} >0,
\]
and is ``seen'' by the slow loop as an additional incoherent source.
This leads to the following expression for image plane intensity and
photon counts,

\begin{alignat*}{1}
I_{i}^{IP} & =\dot{N}_{S}\left\Vert G_{i,slow}^{IP}\bm{\epsilon}_{slow}^{CL}+\mathbf{E}_{0,i}^{WS}\right\Vert ^{2},\\
y_{i}^{IP} & \sim poisson\left(\left(I_{i}^{IP}+D_{ext,i}^{IP}+D_{int,i}^{IP}+D_{jit,i}^{IP}\right)t_{s,slow}\right),
\end{alignat*}
which can then be used to find the contrast bounds per Sec.~\ref{sub:implementation}.

\subsection{\label{sub:RST}Closed-loop Speckles Floor for the RST}

For our final application, we compute bounds on the steady-state speckle
intensity that can be maintained on RST with image plane HOWFS, i.e.,
without periodically pointing at a reference star to recreate the
dark hole (\cite{bailey2018lessons}). Our analysis is based on the publicly
available OS 9 simulation (\cite{krist2020observing}), and it suggests
that the speckles can be maintained continuously
below the dominant detector noise for the parameters of this particular scenario.

We first need to extract the information about the statistical properties of the open loop drifts from the OS9 data post-coronagraph electric fields (instead of the underlying wavefronts, although  starting with wavefront would yield similar results).  To do so, we picked $12$ uninterrupted
sequences of image-plane electric fields (at $0.1\:\lambda/D$ resolution)
and images (at $0.42\:\lambda/D$ resolution) during which the telescope
had a fixed alignment, pointing at a target star. Each such sequence
contains between $K=17$ and $K=21$ electric fields taken $5$ minutes
apart and corresponding to $c_{pol}=4$ polarizations and $c_{wvl}=9$
wavelengths. The electric fields were resampled to the resolution
of the images, and only $N_{pix}=1604$ pixels between $3$ and $10\lambda/D$
were selected, giving the following vector sequences,
\[
\left\{ \mathbf{E}_{k,1}\right\} _{k=1}^{K_{1}},...,\left\{ \mathbf{E}_{k,12}\right\} _{k=1}^{K_{12}}\subset\mathbb{R}^{2c_{pol}c_{wvl}N_{pix}}.
\]

\begin{figure}
\includegraphics[scale=0.75]{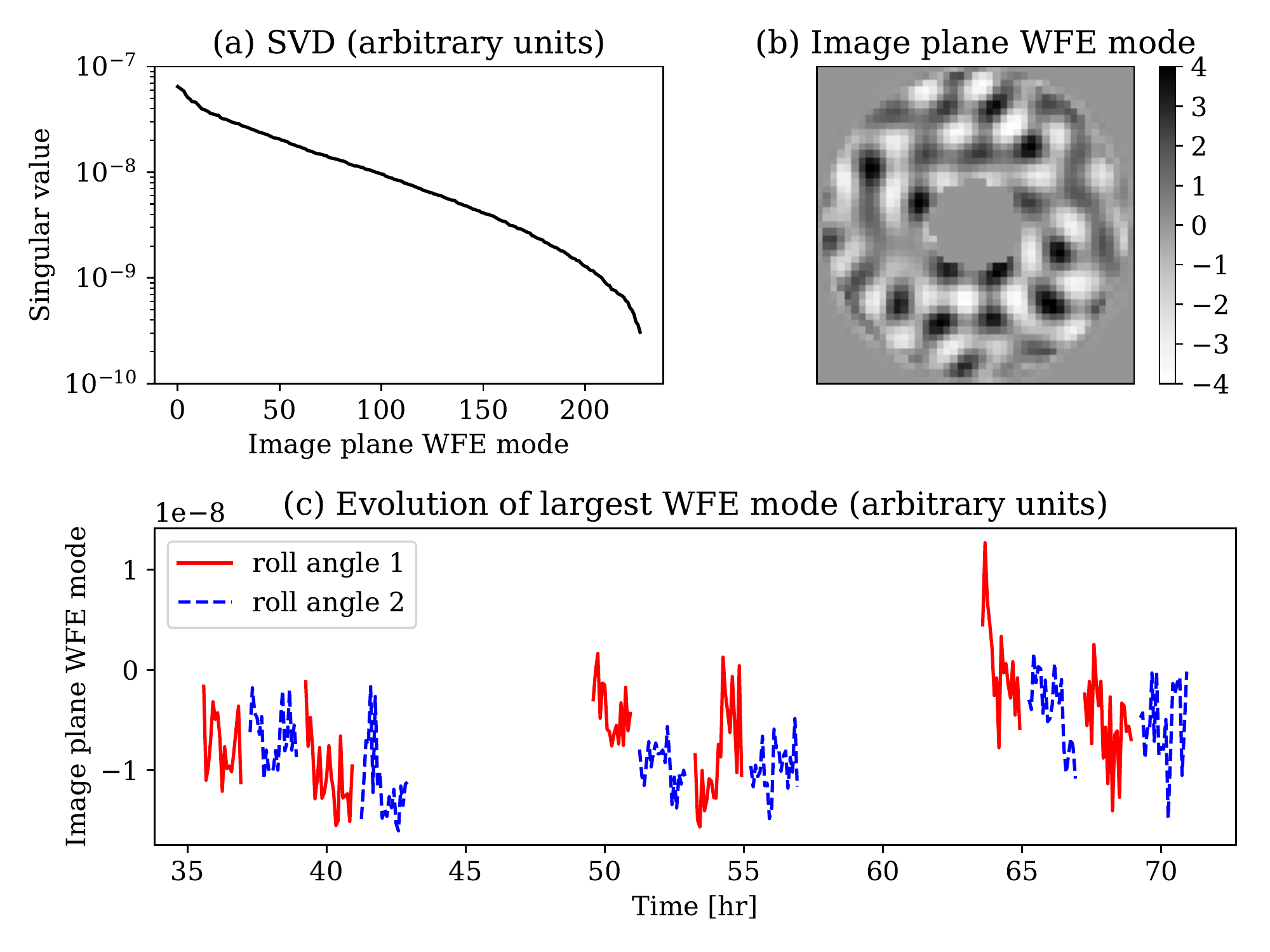}\caption{\label{fig:RST_modes}Singular value decomposition of the image-plane electric
field increments in RST observation scenario 9 (\cite{krist2020observing}).
(a) The singular values. (b) The largest WFE mode (one of the channels).
(c) The evolution of the largest mode (proportional to its contribution to the electric field). Only times at which the telescope
was pointing at the target star are shown, and the sequences are split
according to its roll angle.}
\end{figure}

In order to compute the WFE drift modes from the simulated data, we
arranged the electric field increments into a $115488\times228$ matrix,
\[
\Delta\Upsilon = \begin{bmatrix}\cdots & \mathbf{E}_{k+1,1}-\mathbf{E}_{k,1} & \cdots & \cdots & \mathbf{E}_{k+1,12}-\mathbf{E}_{k,12} & \cdots\end{bmatrix}\in\mathbb{R}^{2c_{pol}c_{wvl}N_{pix}\times r},
\]
where $r=K_{1}-1+...+K_{12}-1=228$ is the number of empirical WFE
modes. Assuming that the modes exhibit Brownian motion, the singular
value decomposition of the increments matrix, $\Delta\Upsilon=U\Sigma V^{T}$,
gives estimates of the WFE sensitivity matrix and drift covariance,

\[
G^{IP}=U\in\mathbb{R}^{2c_{pol}c_{wvl}N_{pix}\times r},\:Q(5\:\mathrm{min})=\frac{1}{r-1}\Sigma^{2}\in\mathbb{R}^{r\times r}.
\]
Figure~\ref{fig:RST_modes}(c) shows the evolution of the largest
mode (Fig.~\ref{fig:RST_modes}(b)) which appears to be neither differentiable, nor discontinuous thus, at least partially, justifying the Brownian-motion in Sec.~\ref{sub:discrete_time}. The static
electric field estimate is found by projecting the dynamics modes
out, i.e.,
\[
\mathbf{E}_{0}=\mathbf{E}_{1,1}-UU^{T}\mathbf{E}_{1,1}
\]
(this estimate depends on the frame number, but the variations between
frames are insignificant in OS 9).

The images in \cite{krist2020observing} correspond to $t_{s}=5\:\mathrm{min}$
long exposures on the target star, 47 UMa. In this scenario, the photon
flux from the star was $\dot{N}_{S}=8.2\cdot10^{7}\:\mathrm{s}^{-1}$
and we estimated the detector noise (i.e., ``incoherent'' flux,
$D_{i}$) to be $1.3$ electrons per exposure at each pixel, based on the images in OS 9. The electric
fields were scaled to give the correct image intensities when squared and
multiplied by $\dot{N}_{S}$.

\begin{figure}
\includegraphics[scale=0.75]{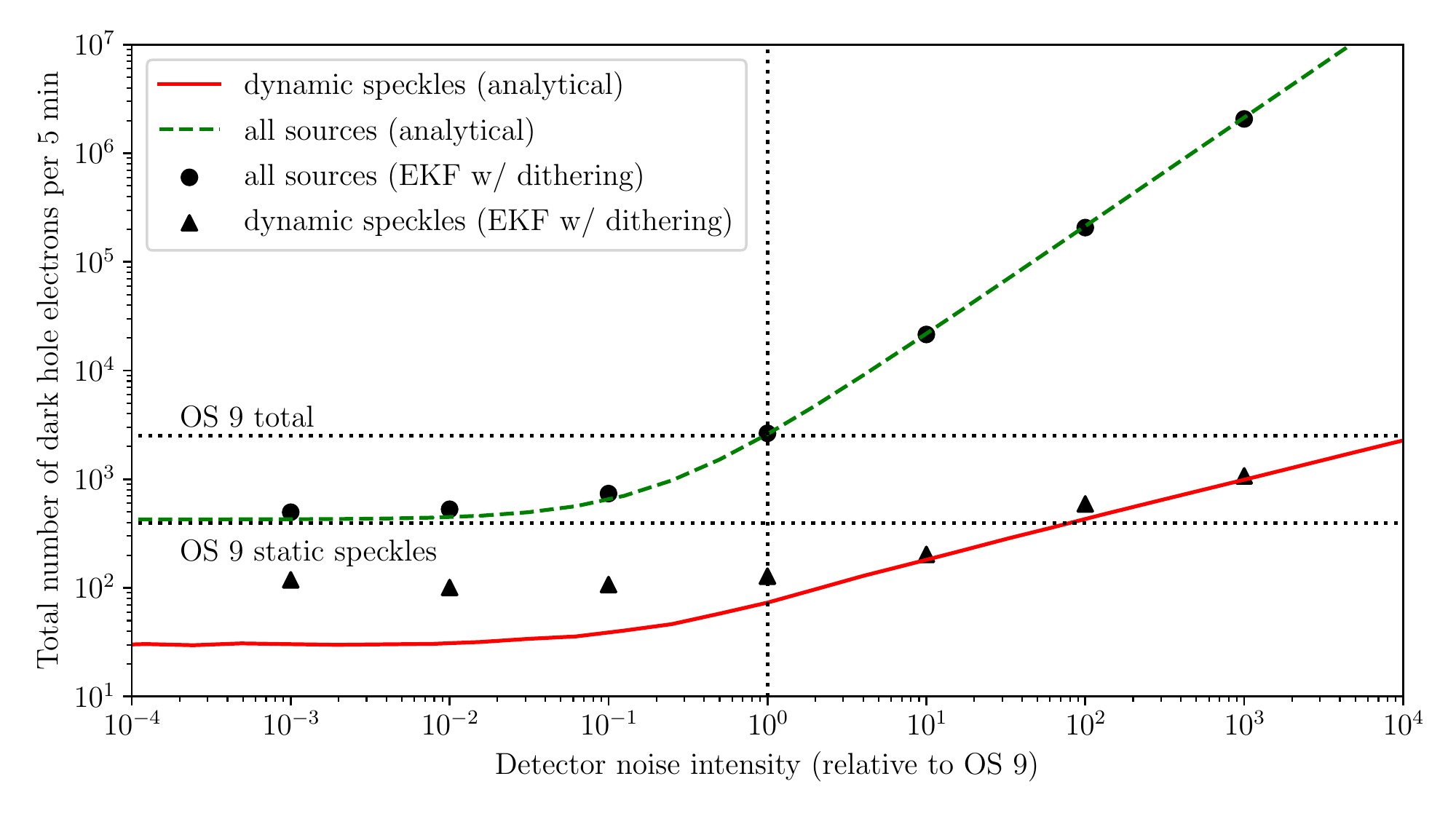}\caption{\label{fig:os9_results}Estimated image-plane intensity as a function
of detector noise, based on RST OS 9 open-loop simulations (horizontal dotted lines), closed-loop
analytical bounds (solid red line and dashed green line) and EKF simulations (triangles and circles). In the given data (vertical dotted line), sources
internal to the telescope (clock-induced charge, etc.) are dominant
in both closed- and open-loop observation. When observing brighter
stars (left side), static speckles become dominant. In any case, closing
the loop (circles and dashed green line) does not significantly impact the intensity and would therefore
be preferable to a lower-duty-cycle open-loop approach.}

\end{figure}

This time, we do not provide any algebraically-derived limits, and compute the closed-loop bounds  via Algorithm~1. For Fig.~\ref{fig:os9_results} we varied the relative contribution
of detector noise, $D_{i}$, to examine its effects on the closed-loop
speckles and the total intensity at the image plane. At the above
mentioned level of $D_{i}=\frac{1.3}{5\:\mathrm{min}}$, the incoherent
sources constituted over $85\%$ of the electrons in the dark hole.
In a hypothetical scenario where a $10$ times brighter target is
observed instead, the majority of the electrons would come from static
speckles (the contrast floor achieved when creating the dark hole).
In that case, the dynamic speckles driven by wavefront instabilities
would be accurately estimated and well constrained. However,
in accordance with Fig.~\ref{fig:cubic}(a) and surrounding discussion,
the variance of the closed loop WFE increases proportionally to the
cubic root of the incoherent intensity. As a result, even if detector
noise is known and uniform in time, it may have an adverse affect on
the \emph{systematic} error in post-processing.

Figure~\ref{fig:os9_results} also shows the closed-loop intensities
obtained by an EKF of the WFE modes (see appendix~\ref{sec:EKF_os9}).
Similarly to the example in Sec.~\ref{sub:example_single}, the qualitative
behavior of the EKF is generally consistent with the analytical bound, although
there is a factor of $3$ discrepancy between the two (in the limit
of low detector noise). We suspect that a better result could be
achieved if the dither, which is necessary for phase diversity, is
optimized by some sophisticated choice of DM actuations. Nevertheless, the intensity remains dominated by incoherent
sources, or static speckles in the limit of negligible detector noise.

We conclude that it is possible, at least in theory, to maintain a
steady contrast throughout the nominal RST observation sequence by closing the loop in the image plane. Changing the orientation of the telescope to periodically observe a reference
star would then become unnecessary. Besides reducing the duty cycle,
such maneuvers might also increase WFE drift rate and jitter residual
due to reaction wheels. A future simulation of an uninterrupted observation
scenario would be necessary to assess the benefits of a closed-loop
approach.

\section{\label{sec:conclusions}Conclusions}

We proposed a method for computing a lower bound on the variance of
post-LOWFS and post-HOWFS wavefront modes. The method yields contrast estimates that reproduce previous theoretical work (\cite{guyon2005limits}) in some bounding cases, generalize it to recursive estimation and non-atmospheric WFE, are consistent with end-to-end AO simulations, and are consistent with dark hole maintenance simulations of the RST based on OS 9. Our analytical approach avoids joint end-to-end simulations of the coronagraph with its wavefront control loops. As a result, the optics need to be propagated just once when computing WFE sensitivity matrices, even when assessing a large number of observation scenarios.

Using this approach, we showed that recursive
estimation that takes into account WFE dynamics gives the best contrast, and derived power laws of their dependencies on photon flux, detector noise and
temporal PSD of the WFE. Based on RST OS 9, we predict that it should
be possible to \emph{continuously}
reject high-order wavefront perturbations due to thermal drift of
the OTA with negligible contrast loss. The analysis of post-processing S/N as a function of the residual wavefront variance is left for future work.

The basic implicit equation for a bound on closed-loop WFE variance is derived
in Sec.~\ref{sub:discrete_time} for when the open-loop
WFE modes exhibit Brownian motion and all noise sources are Poisson-distributed. This bound relies on the average Fisher information contained in sensor photon counts and the Cram{\'e}r-Rao inequality.
In Sec.~\ref{sub:continuous_time},
it is extended to linear dynamics of an arbitrary order and
continuous in time. Two algorithms to approximately compute these bounds
are given in Sec.~\ref{sub:implementation}.

If the WFE drift modes are ``decoupled'' in the sense described
in Sec.~\ref{sub:special_discrete}, it becomes possible to derive
closed form expressions for their residual variance in some special
cases. In particular, it is shown that the best contrast is achieved in the
limit of zero exposure time, and that batch estimation is less
``efficient'' than recursive estimation. When incoherent sources are dominant,
the WFE variance increases proportionally to the cubic root of the
incoherent intensity (a detail which might play a role in post-processing
where the two sources have qualitatively different behaviors). 

In Sec.~\ref{sub:special_AO}, we derive the scaling of closed-loop contrasts with respect to WFE drift magnitude and star brightness under some special assumptions on the open-loop PSD. Our results generalize previous derivations and numerical studies of AO systems, and suggest that currently existing methods do not yet reach theoretical performance limits. Specifically, for WFEs with PSDs that decay rapidly with frequency, the recursive estimation bounds have more favorable scaling laws than both batch estimation and more modern controllers.

Section~\ref{sub:example_single} compares the analytical bounds
to recursive (EKF) and batch (pair probing) estimation algorithms
for a theoretical single-pixel system. Although the bounds are not tight, their qualitative behavior matches simulation results. In Section~\ref{sub:combining_bounds},
we consider a joint analysis of fast LOWFS and much slower HOWFS loops.
The combined bounds can be found by first computing the LOWFS residuals,
which then appear as an incoherent source when computing the final
contrast estimates.

In Sec.~\ref{sub:RST}, based on OS 9, we estimate a bound on the
image plane intensity that could be maintained by RST while continuously
observing the target star, 47 UMa. In this scenario, the dominant
source of electrons are internal to the telescope (i.e., Poisson-distributed dark current). The contributions of dynamic speckles and DM probes necessary for wavefront sensing are less significant. As a result, we conclude that it should, at
least in theory, be possible to observe a dim target star continuously
without periodically switching to a reference star for the purpose
of dark hole maintenance. In our HOWFS numerical simulations, the
error covariances of the Extended Kalman Filter were larger than the
analytical bound by a factor of up to $3$. Since it is necessary, for estimation purposes, to
introduce phase diversity via DM probing or dithering, we speculate
that the proposed lower bound is unattainable.

\appendix

\section{Derivations}

\subsection{\label{sec:finite_exposure}Recursive WFE Covariance for Finite Exposure Time}
Equation~(\ref{eq:bound}) is the key equation that we use throughout the
paper that relates the closed-loop WFE modes covariance, $P+Q$, to
the average information obtained from measurements, $\mathrm{E}_{\bm{\epsilon}^{CL}}\left\{ \left.{\cal I}\right|P+Q\right\} $.
\begin{equation}
P^{-1}-\left(P+Q\right)^{-1}=\mathrm{E}_{\bm{\epsilon}^{CL}}\left\{ \left.{\cal I}\right|P+Q\right\} . \tag{\ref{eq:bound}}
\end{equation}
It implicitly approximates the WFE modes as fixed throughout the exposure,
and equal to their value at the end of the exposure. In practice,
the WFE covariance increases linearly from $P$ at the beginning of
the exposure, to $P+Q$ at the end, giving a time averaged covarinace
of $P+\frac{1}{2}Q$. Moreover, fluxes also vary in time throughout
the exposure, making the co-added photon counts less indicative of the
flux at the end.

Here, for completeness, we provide a more subtle analysis that takes
WFE drift during the exposure into account. It results in the follow
relation for $P$
\begin{equation}
P=P+Q-\left(P+\frac{1}{2}Q\right)\left(P+\mathrm{E}_{\bm{\epsilon}^{CL}}\left\{ \left.{\cal I}^{-1}\right|P+\frac{1}{2}Q\right\} +\frac{1}{3}Q\right)^{-1}\left(P+\frac{1}{2}Q\right),\label{eq:finite_exposure_P}
\end{equation}
and can be used in Algorithm 1 instead of the less precise Eq.~(\ref{eq:bound}).
Additionally, instead of the average contrast at the end of the exposure,
Eq.~(\ref{eq:contrast}), one can measure performance based on the average contrast
throughout,
\begin{equation}
C=C_{0}+\underset{i}{\sum}\left[\frac{D_{i}^{IP}}{\dot{N}_{S}^{IP}}+\mathrm{trace}\left\{ G_{i}^{IP}\left(P+\frac{1}{2}Q\right)\left(G_{i}^{IP}\right)^{T}\right\} \right].\label{eq:finite_exposure_C}
\end{equation}

While these expressions are more accurate, they are also cumbersome
and less intuitive. They give the same estimates as Eqs.~(\ref{eq:bound}) and (\ref{eq:contrast})
in the limit of short exposure time $t_{s}\rightarrow0$ where ${\cal I}^{-1}\gg P\gg Q$,
and differ only slightly in the limit $t_{s}\rightarrow\infty$ (as we show
 after the derivation below). For these reasons, we use the simpler
expressions throughout the paper.

We now derive Eq.~(\ref{eq:finite_exposure_P}) by splitting the
$k$-th exposure into $M$ sub intervals in which the wavefront errors accumulate
in open-loop. The coefficients of the WFE modes corresponding to times
$\left(k+\frac{m}{M}\right)t_{s}$ are denoted as $\bm{\epsilon}_{k+\frac{m}{M}}^{OL}$ and related via
\[
\bm{\epsilon}_{k+\frac{m+1}{M}}^{OL}=\bm{\epsilon}_{k+\frac{m}{M}}^{OL}+\bm{v}_{k+\frac{m}{M}},\:\bm{v}_{k+\frac{m}{M}}\sim{\cal N}\left(\bm{0},\frac{1}{M}Q\right).
\]
Note that $\bm{\epsilon}_{k+0}^{OL}=\bm{\epsilon}_{k}^{CL}\sim{\cal N}\left(\bm{0},P_{k}\right)$.

These coefficients are each ``sensed'' by the wavefront sensors
and then averaged,
\[
\hat{\bm{\epsilon}}_{k+1,batch}^{OL}=\frac{1}{M}\underset{m=1}{\overset{M}{\sum}}\hat{\bm{\epsilon}}_{k+\frac{m}{M},batch}^{OL}.
\]
Here, $\hat{\bm{\epsilon}}_{k+1,batch}^{CL}$ is the output of the
wavefront sensor at the end of the exposure and
\[
\hat{\bm{\epsilon}}_{k+\frac{m}{M},batch}^{OL}=\bm{\epsilon}_{k+\frac{m}{M}}^{OL}+\bm{w}_{k+\frac{m}{M}}
\]
are hypothetical estimates based on photon counts during the short $\frac{1}{M}t_s$ intervals. We assume that the noise $\bm{w}_{k+\frac{m}{M}}$ is zero-mean and
normally distributed with covariance $\left(\frac{t_{s}}{M}\dot{{\cal I}}(\bm{\epsilon}_{k+\frac{m}{M}}^{CL})\right)^{-1}$
where $\dot{{\cal I}}$ is the information rate based on Eq.~(\ref{eq:information}).

We now wish to find the recursive estimate, $\hat{\bm{\epsilon}}_{k+1}^{OL}$,
which takes into account both the measurement, $\hat{\bm{\epsilon}}_{k+1,batch}^{OL}$,
and the priors $\bm{\epsilon}_{k}^{CL}\sim{\cal N}\left(\bm{0},P_{k}\right)$
and $\bm{\epsilon}_{k+1}^{OL}-\bm{\epsilon}_{k}^{CL}\sim{\cal N}\left(\bm{0},Q\right)$.
First, note that
\begin{equation}
\hat{\bm{\epsilon}}_{k+1,batch}^{OL}=\frac{1}{M}\underset{m=1}{\overset{M}{\sum}}\left(\bm{\epsilon}_{k+\frac{m}{M}}^{OL}+\bm{w}_{k+\frac{m}{M}}\right)=\bm{\epsilon}_{k}^{CL}+\underset{m=1}{\overset{M}{\sum}}\frac{M-m+1}{M}\bm{v}_{k+\frac{m}{M}}+\frac{1}{M}\underset{l=1}{\overset{M}{\sum}}\bm{w}_{k+\frac{m}{M}},\label{eq:finite_batch_est}
\end{equation}
and hence $\hat{\bm{\epsilon}}_{k+1,batch}^{OL}$ is a sum of independent normally-distributed variables. Its total
covariance is
\[
\mathrm{cov}\hat{\bm{\epsilon}}_{k+1,batch}^{OL}=\Sigma_{M}=P_{k}+\underset{m=1}{\overset{M}{\sum}}\frac{(M-m+1)^{2}}{M^{3}}Q+\frac{1}{M}\underset{m=1}{\overset{M}{\sum}}\left(t_{s}\dot{{\cal I}}(\bm{\epsilon}_{k+\frac{m}{M}}^{CL})\right)^{-1}.
\]

It can be shown that the \emph{a-posteriori }maximum-likelihood estimates
of all of the above variables are given by
\begin{alignat*}{1}
\hat{\bm{\epsilon}}_{k|k+1}^{CL}= & P_{k}\Sigma_{M}^{-1}\hat{\bm{\epsilon}}_{k+1,batch}^{OL},\\
\hat{\bm{v}}_{k+\frac{m}{M}|k+1}= & \frac{M-m+1}{M^{2}}Q\Sigma_{M}^{-1}\hat{\bm{\epsilon}}_{k+1,batch}^{OL},\\
\hat{\bm{w}}_{k+\frac{m}{M}}= & \left(t_{s}\dot{{\cal I}}(\bm{\epsilon}_{k+\frac{m}{M}}^{CL})\right)^{-1}\Sigma_{M}^{-1}\hat{\bm{\epsilon}}_{k+1,batch}^{OL}.
\end{alignat*}
Therefore, the \emph{a-posteriori }maximum-likelihood estimate of
$\hat{\bm{\epsilon}}_{k+1}^{OL}=\hat{\bm{\epsilon}}_{k+1|k+1}^{OL}$
is
\[
\hat{\bm{\epsilon}}_{k+1}^{OL}=\hat{\bm{\epsilon}}_{k|k+1}^{CL}+\underset{m=1}{\overset{M}{\sum}}\hat{\bm{v}}_{k+\frac{m}{M}|k+1}=\left(P_{k}+\underset{m=1}{\overset{M}{\sum}}\frac{M-m+1}{M^{2}}Q\right)\Sigma_{M}^{-1}\hat{\bm{\epsilon}}_{k+1,batch}^{OL}.
\]
In the limit $M\rightarrow\infty$, we have $\underset{m=1}{\overset{M}{\sum}}\frac{M-m+1}{M^{2}}\rightarrow\frac{1}{2}$
and $\underset{m=1}{\overset{M}{\sum}}\frac{(M-m+1)^{2}}{M^{3}}\rightarrow\frac{1}{3}$,
hence
\[
\hat{\bm{\epsilon}}_{k+1}^{OL}\approx\left(P_{k}+\frac{1}{2}Q\right)\left(P_{k}+\mathrm{E}_{\bm{\epsilon}^{CL}}\left\{ \left.{\cal I}^{-1}\right|P_{k}+\frac{1}{2}Q\right\} +\frac{1}{3}Q\right)^{-1}\hat{\bm{\epsilon}}_{k+1,batch}^{OL},
\]
where we replaced $\frac{1}{M}\underset{m=1}{\overset{M}{\sum}}\left(t_{s}\dot{{\cal I}}(\bm{\epsilon}_{k+\frac{m}{M}}^{CL})\right)^{-1}$
with its approximate value in the middle of the exposure, $\mathrm{E}_{\bm{\epsilon}^{CL}}\left\{ \left.{\cal I}^{-1}\right|P+\frac{1}{2}Q\right\} $.

To compute the closed-loop covariance at the beginning of the $k+1$
exposure, $P_{k+1}=\mathrm{cov}\bm{\epsilon}_{k+1}^{CL}$, note that
\begin{gather*}
\bm{\epsilon}_{k+1}^{CL}=\hat{\bm{\epsilon}}_{k+1}^{OL}-\bm{\epsilon}_{k+1}^{OL}=\\
=\left(P_{k}+\frac{1}{2}Q\right)\left(P_{k}+\mathrm{E}_{\bm{\epsilon}^{CL}}\left\{ \left.{\cal I}^{-1}\right|P_{k}+\frac{1}{2}Q\right\} +\frac{1}{3}Q\right)^{-1}\hat{\bm{\epsilon}}_{k+1,batch}^{OL}-\bm{\epsilon}_{k+1}^{OL}.
\end{gather*}
Using the expression for $\hat{\bm{\epsilon}}_{k+1,batch}^{OL}$ in
Eq.~(\ref{eq:finite_batch_est}) and $\bm{\epsilon}_{k+1}^{OL}=\bm{\epsilon}_{k}^{CL}+\underset{m=1}{\overset{M}{\sum}}\bm{v}_{k+\frac{m}{M}}$,
one can derive the expression for $P_{k+1}$,
\[
P_{k+1}=P_{k}+Q-\left(P_{k}+\frac{1}{2}Q\right)\left(P_{k}+\mathrm{E}_{\bm{\epsilon}^{CL}}\left\{ \left.{\cal I}^{-1}\right|P_{k}+\frac{1}{2}Q\right\} +\frac{1}{3}Q\right)^{-1}\left(P_{k}+\frac{1}{2}Q\right).
\]
Equation~(\ref{eq:finite_exposure_P}) then follows as the steady state case limit, $P_{k+1}=P_{k}=P$.

We now compare Eqs.~(\ref{eq:finite_exposure_P}) and (\ref{eq:finite_exposure_C})
to (\ref{eq:bound}) and (\ref{eq:contrast}) under the assumptions of Sec~\ref{sub:special_discrete}. The information, contrast and covariance expressions in Eqs.~(\ref{eq:diagonalized_information})-(\ref{eq:diagonalized_contrast})
become
\begin{alignat*}{1}
{\cal I}_{j}\approx & 4\dot{N}_{S}t_{s}\frac{\underset{l=1}{\overset{r}{\sum}}(p_{l}^{2}+\frac{1}{2}q_{l}^{2})\Lambda_{l}^{2}+\frac{1}{2}\left\Vert \mathbf{E}_{0}\right\Vert ^{2}}{2\underset{l=1}{\overset{r}{\sum}}(p_{l}^{2}+\frac{1}{2}q_{l}^{2})\Lambda_{l}^{2}+\left\Vert \mathbf{E}_{0}\right\Vert ^{2}+\dot{N}_{S}^{-1}D}\Lambda_{j}^{2},\\
p_{j}^{2}\approx & p_{j}^{2}+q_{j}^{2}-\frac{\left(p_{j}^{2}+\frac{1}{2}q_{j}^{2}\right)^{2}}{p_{j}^{2}+{\cal I}_{j}^{-1}+\frac{1}{3}q_{j}^{2}},\\
C= & C_{0}+2\underset{j}{\sum}\left(p_{j}^{2}+\frac{1}{2}q_{j}^{2}\right)\left(\Lambda_{j}^{IP}\right)^{2}+\frac{D^{IP}}{\dot{N}_{S}^{IP}}.
\end{alignat*}

In the case of short exposure time we have
\[
q_{j}^{2}=\xi_{j}^{2}t_{s}\ll p_{j}^{2}\ll t_{s}^{-1}{\cal \dot{I}}_{j}^{-1}={\cal I}_{j}^{-1},
\]
as $t_{s}$ becomes small. Then, ${\cal I}_{j}$ and $C$ do not explicitly
depend on $q_{j}^{2}$ and their expressions above become identical to
Eqs.~(\ref{eq:diagonalized_information}) and (\ref{eq:diagonalized_contrast}). Equation~(\ref{eq:diagonalized_bound}) also converges to its finite-exposure
equivalent since,
\[
p_{j}^{-2}-(p_{j}^{2}+\xi_{j}^{2}t)^{-1}\approx p_{j}^{-2}-p_{j}^{-2}(1-p_{j}^{-2}\xi_{j}^{2}t)=p_{j}^{-4}\xi_{j}^{2}t
\]
and
\[
\frac{\left(p_{j}^{2}+\frac{1}{2}\xi_{j}^{2}t_{s}\right)^{2}}{p_{j}^{2}+{\cal I}_{j}^{-1}+\frac{1}{3}\xi_{j}^{2}t_{s}}\approx{\cal I}_{j}p_{j}^{4}.
\]
We conclude that the two approaches give the same bounds at the short-exposure limit which is where the recursive estimator is optimal.

In the case of long exposure time, $\dot{N}_{S}^{-1}D\ll\xi_{j}^{2}t_{s}$
as $t_{s}$ becomes large. We have ${\cal I}_{j}\approx2\dot{N}_{S}t_{s}\Lambda_{j}^{2}$
and can solve for $p_{j}$ and $C$,
\begin{alignat*}{1}
p_{j}^{2}= & q_{j}^{2}\sqrt{\frac{1}{12}+\frac{1}{2\dot{N}_{S}t_{s}\Lambda_{j}^{2}q_{j}^{2}}}\approx\sqrt{\frac{1}{12}}q_{j}^{2},\\
C-C_{0}= & \underset{j}{\sum}\left(\sqrt{\frac{1}{3}+\frac{2}{\dot{N}_{S}t_{s}\Lambda_{j}^{2}q_{j}^{2}}}+1\right)\left(\Lambda_{j}^{IP}q_{j}\right)^{2}\approx\left(1+\sqrt{\frac{1}{3}}\right)\underset{j}{\sum}\left(\Lambda_{j}^{IP}q_{j}\right)^{2}.
\end{alignat*}
Note that the contrast loss is smaller by a factor of about $1.3$
than the one obtained from Eq.~(\ref{eq:diagonalized_contrast_no_incoherent}),
\[
C-C_{0}=\underset{j}{\sum}\left(\sqrt{1+\frac{2}{\dot{N}_{S}t_{s}\Lambda_{j}^{2}q_{j}^{2}}}+1\right)\left(\Lambda_{j}^{IP}q_{j}\right)^{2}\approx2\underset{j}{\sum}\left(\Lambda_{j}^{IP}q_{j}\right)^{2}.
\]

\subsection{\label{sec:optimality}Optimality of Zero Exposure Time in the Presence of Poisson Distributed Noise Sources}
We begin with the assumptions in Sec.~\ref{sub:special_discrete} and wish to prove that the contrast $C$ given by Eq.~(\ref{eq:diagonalized_contrast})
 achieves its infimum w.r.t. exposure time $t_{s}$ at the limit $t_{s}=0$. In particular, we assume that all intensity sources are Poisson distributed and that the wavefront modes drift independently via Brownian motion ($q_{j}=\xi_{j}^{2}t_{s}$). We will only show that the one-sided derivative of the contrast at $t_s=0$ is positive,
\[
\frac{\partial C}{\partial t_{s}}=2\underset{j}{\sum}\left(\Lambda_{j}^{IP}\right)^{2}\left(\left.\frac{\partial p_{j}^{2}}{\partial t_{s}}\right|_{t_{s}=0}+\xi_{j}^{2}\right)>0.
\]
%\[
%\frac{\partial C}{\partial t_{s}}=\underset{j}{\sum}\left(\Lambda_{j}^{IP}\right)^{2}\left(\left.2\frac{\partial p_{j}^{2}}{\partial t_{s}}\right|_{t_{s}=0}+\xi_{j}^{2}\right)=0.
%\]

Combining Eqs.~(\ref{eq:diagonalized_information}) and~(\ref{eq:diagonalized_bound}),
we get
\begin{equation}
p_{j}^{-2}-(p_{j}^{2}+\xi_{j}^{2}t_{s})^{-1}=4\dot{N}_{S}t_{s}\frac{\underset{l=1}{\overset{r}{\sum}}(p_{l}^{2}+\xi_{l}^{2}t_{s})\Lambda_{l}^{2}+\frac{1}{2}\left\Vert \mathbf{E}_{0}\right\Vert ^{2}}{2\underset{l=1}{\overset{r}{\sum}}(p_{l}^{2}+\xi_{l}^{2}t_{s})\Lambda_{l}^{2}+\left\Vert \mathbf{E}_{0}\right\Vert ^{2}+\dot{N}_{S}^{-1}D}\Lambda_{j}^{2},\label{eq:diagonalized_information_bound}
\end{equation}
which reduces to Eq.~(\ref{eq:diagonal_zero_t}) in the limit $t_{s}\rightarrow0$.
Following the discussion and definitions below Eq.~(\ref{eq:diagonal_zero_t}),
we rewrite 
\begin{alignat*}{1}
\left\Vert \mathbf{E}_{0}\right\Vert ^{2} & =\frac{\sqrt{2}\underset{l}{\sum}\xi_{l}\Lambda_{l}}{\sqrt{\dot{N}_{s}}}\sigma_{0},\\
D= & \sqrt{2\dot{N}_{s}}\underset{l}{\sum}\xi_{l}\Lambda_{l}\delta,
\end{alignat*}
and expand $p_{j}^{2}$ about its solution at $t_{s}=0$,
\[
p_{j}^{2}=\frac{1}{\sqrt{2\dot{N}_{S}}}\frac{\xi_{j}}{\Lambda_{j}}\bar{p}^{2}+\frac{\partial p_{j}^{2}}{\partial t_{s}}t_{s}+o(t_s).
\]

Keeping terms up to first order, Eq.~(\ref{eq:diagonalized_information_bound}) becomes
\begin{gather*}
\left(\frac{1}{\sqrt{2\dot{N}_{S}}}\frac{\xi_{j}}{\Lambda_{j}}\bar{p}^{2}+\frac{\partial p_{j}^{2}}{\partial t_{s}}t_{s}\right)^{-1}-\left(\frac{1}{\sqrt{2\dot{N}_{S}}}\frac{\xi_{j}}{\Lambda_{j}}\bar{p}^{2}+\frac{\partial p_{j}^{2}}{\partial t_{s}}t_{s}+\xi_{j}^{2}t_{s}\right)^{-1}=\\
=2\dot{N}_{S}t_{s}\frac{2\underset{l}{\sum}\left(\Lambda_{l}^{2}\frac{\partial p_{l}^{2}}{\partial t_{s}}+\xi_{l}^{2}\Lambda_{l}^{2}\right)\sqrt{\dot{N}_{S}}t_{s}+\sqrt{2}\underset{l}{\sum}\xi_{l}\Lambda_{l}\left(\bar{p}^{2}+\sigma_{0}\right)}{2\underset{l}{\sum}\left(\Lambda_{l}^{2}\frac{\partial p_{l}^{2}}{\partial t_{s}}+\xi_{l}^{2}\Lambda_{l}^{2}\right)\sqrt{\dot{N}_{S}}t_{s}+\sqrt{2}\underset{l}{\sum}\xi_{l}\Lambda_{l}\left(\bar{p}^{2}+\sigma_{0}+\delta\right)}\Lambda_{j}^{2}
\end{gather*}
and after some algebra it can be written as
\begin{gather*}
2\underset{l}{\sum}\Lambda_{l}^{2}\left(\frac{\partial p_{l}^{2}}{\partial t_{s}}+\xi_{l}^{2}\right)\sqrt{\dot{N}_{S}}t_{s}+\sqrt{2}\underset{l}{\sum}\xi_{l}\Lambda_{l}\left(\bar{p}^{2}+\sigma_{0}+\delta\right)=\\
=2\dot{N}_{S}\left(2\underset{l}{\sum}\Lambda_{l}^{2}\left(\frac{\partial p_{l}^{2}}{\partial t_{s}}+\xi_{l}^{2}\right)\sqrt{\dot{N}_{S}}t_{s}+\sqrt{2}\underset{l}{\sum}\xi_{l}\Lambda_{l}\left(\bar{p}^{2}+\sigma_{0}\right)\right)\left(\frac{1}{\sqrt{2\dot{N}_{S}}}\bar{p}^{2}+\frac{\Lambda_{j}}{\xi_{j}}\frac{\partial p_{j}^{2}}{\partial t_{s}}t_{s}+\xi_{j}\Lambda_{j}t_{s}\right)\left(\frac{1}{\sqrt{2\dot{N}_{S}}}\bar{p}^{2}+\frac{\Lambda_{j}}{\xi_{j}}\frac{\partial p_{j}^{2}}{\partial t_{s}}t_{s}\right).
\end{gather*}
The coefficients of the first power of $t_{s}$ must be equal on both
sides, i.e.,
\begin{gather*}
2\underset{l}{\sum}\Lambda_{l}^{2}\left(\frac{\partial p_{l}^{2}}{\partial t_{s}}+\xi_{l}^{2}\right)\sqrt{\dot{N}_{S}}=\\
=2\underset{l}{\sum}\Lambda_{l}^{2}\left(\frac{\partial p_{l}^{2}}{\partial t_{s}}+\xi_{l}^{2}\right)\sqrt{\dot{N}_{S}}\bar{p}^{4}+2\sqrt{\dot{N}_{S}}\left(\underset{l}{\sum}\xi_{l}\Lambda_{l}\right)\left(\bar{p}^{2}+\sigma_{0}\right)\bar{p}^{2}\left(2\frac{\Lambda_{j}}{\xi_{j}}\frac{\partial p_{j}^{2}}{\partial t_{s}}+\xi_{j}\Lambda_{j}\right).
\end{gather*}
After slight rearrangement, we have
\begin{equation}
(1-\bar{p}^{4})\underset{l}{\sum}\Lambda_{l}^{2}\left(\frac{\partial p_{l}^{2}}{\partial t_{s}}+\xi_{l}^{2}\right)=\left(\underset{l}{\sum}\xi_{l}\Lambda_{l}\right)\left(\bar{p}^{2}+\sigma_{0}\right)\bar{p}^{2}\frac{\Lambda_{j}}{\xi_{j}}\left(2\frac{\partial p_{j}^{2}}{\partial t_{s}}+\xi_{j}^{2}\right),\:\forall j.\label{eq:optimality_proof}
\end{equation}

%Note that $\bar{p}^{4}>1$ since
%$\bar{p}^{2}$ is a solution of Eq.~(\ref{eq:p_bar_equation}),
%\[
%\left(\bar{p}^{2}+\sigma_{0}\right)\left(\bar{p}^{4}-1\right)=\delta>0.
%\]
%Therefore, each $2\frac{\partial p_{j}^{2}}{\partial t_{s}}+\xi_{j}^{2}$ on the right hand side of Eq.~(\ref{eq:optimality_proof}) has the opposite sign of ${\sum}\Lambda_{l}^{2}\left(2\frac{\partial p_{l}^{2}}{\partial t_{s}}+\xi_{l}^{2}\right)$ on the left hand side. This can only be true when $2\frac{\partial p_{j}^{2}}{\partial t_{s}}+\xi_{j}^{2}=0$ thus $\frac{\partial C}{\partial t_{s}}=0$ at $t_s=0$.

We will now conclude our proof by assuming the opposite of our
claim, i.e., that $\frac{1}{2}\frac{\partial C}{\partial t_{s}}=\underset{l}{\sum}\Lambda_{l}^{2}\left(\frac{\partial p_{l}^{2}}{\partial t_{s}}+\xi_{l}^{2}\right)\le0$,
and reaching a contradiction. First, note that $\bar{p}^{4}>1$ since
$\bar{p}^{2}$ is a solution of Eq.~(\ref{eq:p_bar_equation}),
\[
\left(\bar{p}^{2}+\sigma_{0}\right)\left(\bar{p}^{4}-1\right)=\delta>0.
\]
Then, the assumption $\frac{\partial C}{\partial t_{s}}\le0$ 
leads to
\[
(1-\bar{p}^{4})\underset{l}{\sum}\Lambda_{l}^{2}\left(\frac{\partial p_{l}^{2}}{\partial t_{s}}+\xi_{l}^{2}\right)\ge0,
\]
and therefore the right-hand side of eq.~(\ref{eq:optimality_proof})
is also non-negative. It follows that $2\frac{\partial p_{j}^{2}}{\partial t_{s}}+\xi_{j}^{2}\ge0$
or $\frac{\partial p_{j}^{2}}{\partial t_{s}}\ge-\frac{1}{2}\xi_{j}^{2}$ for all $j$
and thus
\[
0\ge\frac{1}{2}\frac{\partial C}{\partial t_{s}}=\underset{l}{\sum}\Lambda_{l}^{2}\left(\frac{\partial p_{l}^{2}}{\partial t_{s}}+\xi_{l}^{2}\right)\ge\underset{l}{\sum}\Lambda_{l}^{2}\left(-\frac{1}{2}\xi_{l}^{2}+\xi_{l}^{2}\right)>0
\]
which is the desired contradiction.

\subsection{\label{sub:SI_PSD}Closed-loop single WFE Mode Variance with a Simple Integrator}

To derive the performance of the simple integrator, we denote the
transfer function corresponding to Eq.~(\ref{eq:drift_state_equation})
as
\[
\frac{\epsilon}{v}(s)=\begin{bmatrix}1 & 0 & \cdots & 0\end{bmatrix}\left(sI-A\right)^{-1}B=-\frac{\theta(-f_{0})^{\gamma}}{(s+f_{0})^{\gamma}},
\]
where $\epsilon$ is a single ($r=c=1$) open-loop WFE mode and $v$
is white noise. In AO, the deviation of WFS measurement from 
its nominal value, $\Delta\dot{y}$, is small and approximately linear
in the closed loop WFE, $\epsilon^{CL}\in\mathbb{R}$. In other words,
we assume that $\left\Vert G_{i}\mathbf{\epsilon}^{CL}\right\Vert \ll\left\Vert \mathbf{E}_{0,i}\right\Vert $
and that the transfer function between the WFE and the measurement
is a constant,
\[
\frac{\Delta\dot{y}}{\epsilon^{CL}}(s)=\frac{\partial\underset{i}{\sum}I_{i}}{\partial\mathbf{\epsilon}^{CL}}=2\dot{N}_{S}^{WS}\underset{i}{\sum}\left(\mathbf{E}_{0,i}^{WS}\right)^{T}G_{i}^{WS}.
\]
Additionally, the shot noise of constant magnitude depends on the intensity
at the WFS (assuming a perfect intensity source) which can be stated as,
\[
\frac{\Delta\dot{y}}{w}(s)=\sqrt{\dot{N}_{S}^{WS}\underset{i}{\sum}\left(\mathbf{E}_{0,i}^{WS}\right)^{T}\mathbf{E}_{0,i}^{WS}},
\]
where $w$ is also white noise with $\underset{t}{\overset{t+\Delta t}{\int}}wdt\sim{\cal N}\left(0,\Delta t\right)$.
The above transfer functions are a special case of the AO loop presented
in \cite{males2018predictive} without WFS and DM delays.

The control law
\[
\epsilon^{CL}=\epsilon-\hat{\epsilon}
\]
is specified (up to an initial condition) via the transfer function of the estimator,
$\frac{\hat{\epsilon}}{\Delta \dot{y}}(s)$, and yields the following
closed-loop WFE dependency on open-loop WFE and shot noise, 
\[
\epsilon^{CL}=\frac{\frac{\epsilon}{v}(s)}{1+\frac{\hat{\epsilon}}{\Delta\dot{y}}(s)\frac{\Delta\dot{y}}{\epsilon^{CL}}(s)}v-\frac{\frac{\hat{\epsilon}}{\Delta\dot{y}}(s)\frac{\Delta\dot{y}}{w}(s)}{1+\frac{\hat{\epsilon}}{\Delta\dot{y}}(s)\frac{\Delta\dot{y}}{\epsilon^{CL}}(s)}w.
\]
Since $v,w$ are independent white noise, the closed-loop PSD is given
by,
\[
\mathrm{PSD}^{CL}(f)=\left|\frac{\frac{\epsilon}{v}(f)}{1+\frac{\hat{\epsilon}}{\Delta\dot{y}}(f)\frac{\Delta\dot{y}}{\epsilon^{CL}}(f)}\right|^{2}+\left|\frac{\frac{\hat{\epsilon}}{\Delta\dot{y}}(f)\frac{\Delta\dot{y}}{w}(f)}{1+\frac{\hat{\epsilon}}{\Delta\dot{y}}(f)\frac{\Delta\dot{y}}{\epsilon^{CL}}(f)}\right|^{2}
\]
and the variance and contrast contribution by
\begin{alignat*}{1}
\mathrm{var}\left\{ \epsilon^{CL}\right\} = & \underset{0}{\overset{\infty}{\int}}\mathrm{PSD}^{CL}(f)df.\\
\Delta C= & 2\mathrm{var}\left\{ \epsilon^{CL}\right\} \left(\Lambda^{IP}\right)^{2}
\end{alignat*}

In the case of a simple integrator control/estimation law parameterized
by $f_{SI}$,
\[
\frac{\hat{\epsilon}}{\Delta\dot{y}}(s)=\frac{f_{SI}}{2\dot{N}_{S}^{WS}\underset{i}{\sum}\left(\mathbf{E}_{0,i}^{WS}\right)^{T}G_{i}^{WS}}s^{-1},
\]
the variance is
\begin{equation}
\mathrm{var}\left\{ \epsilon^{CL,SI}\right\} \approx\theta^{2}\underset{0}{\overset{\infty}{\int}}\frac{f_{0}^{2\gamma}f^{2}}{(f+f_{SI})^{2}(f+f_{0})^{2\gamma}}df+\dot{{\cal I}}^{-1}\underset{0}{\overset{\infty}{\int}}\frac{f_{SI}^{2}}{(f+f_{SI})^{2}}df,\label{eq:SI_variance}
\end{equation}
where we made the approximation
\[
\dot{{\cal I}}\approx\frac{4\dot{N}_{S}^{WS}\left(\underset{i}{\sum}\left(\mathbf{E}_{0,i}^{WS}\right)^{T}G_{i}^{WS}\right)^{2}}{\underset{i}{\sum}\left(\mathbf{E}_{0,i}^{WS}\right)^{T}\mathbf{E}_{0,i}^{WS}},
\]
by switching the order of summation and division as we did in Eq.~(\ref{eq:diagonalized_information}).
Again we constrain the discussion to the ``pure-integrator'' regime,
$\theta^{2}\dot{{\cal I}}\gg1$, for which the following limit is
applicable,

\[
\underset{f_{0}\rightarrow0}{\lim}\theta^{2}f_{0}{}^{2\gamma}\underset{0}{\overset{\infty}{\int}}\frac{f^{2}}{(f+f_{SI})^{2}(f+f_{0})^{2\gamma}}df\sim\begin{cases}
\frac{\theta^{2}f_{0}^{2}}{f_{SI}} & \gamma=1\\
\frac{\theta^{2}f_{0}^{3}}{f_{SI}^{2}} & \gamma\ge2
\end{cases}.
\]
The variance in Eq.~(\ref{eq:SI_variance}) can be optimized w.r.t.
$f_{SI}$ resulting in
\[
\underset{f_{SI}}{\min}\mathrm{var}\left\{ \epsilon^{CL,SI}\right\} \sim\begin{cases}
\left(\theta^{2}f_{0}^{2}\dot{{\cal I}}^{-1}\right)^{\frac{1}{2}} & \gamma=1\\
\left(\theta^{2}f_{0}^{3}\dot{{\cal I}}^{-2}\right)^{\frac{1}{3}} & \gamma\ge2
\end{cases},
\]
up to some constant. In terms of contrast and the normalized quantities defined in Eq.~(\ref{eq:normalized_PSD}), this gives Eq.~(\ref{eq:SI_scaling}).

\subsection{\label{sec:EKF_os9}EKF of OS 9 WFE Modes}

We detail the EKF corresponding to the WFE dynamics used in
Sec.~\ref{sub:discrete_time} to compute closed-loop intensity estimates
in Fig.~\ref{fig:os9_results}. Similarly
to \cite{pogorelyuk2019dark} we approximate the measurement equation (i.e, Eq.~(\ref{eq:poisson}))
with a normal distribution,
\[
y_{i}\sim{\cal N}\left(\left(I_{i}+D_{i}\right)t_{s},\left(I_{i}+D_{i}\right)t_{s}\right).
\]
In vector notation,
\begin{alignat*}{1}
G & =\begin{bmatrix}\vdots\\
G_{i}\\
\vdots
\end{bmatrix}\in\mathbb{R}^{2cN_{pix}\times r},\\
M & =\begin{bmatrix}\ddots\\
 & \mathbf{1}_{2c}\\
 &  & \ddots
\end{bmatrix}\in\mathbb{R}^{N_{pix}\times2cN_{pix}},\\
\mathbf{D} & =\begin{bmatrix}\vdots\\
D_{i}\\
\vdots
\end{bmatrix}\in\mathbb{R}^{N_{pix}},\\
\mathbf{I}_{k} & =\dot{N}_{S}M\cdot\left(G\left(\bm{\epsilon}_{k}+\mathbf{u}_{k}\right)+\mathbf{E}_{0}\right)^{\circ2}\in\mathbb{R}^{N_{pix}},\\
\mathbf{y}_{k} & \sim{\cal N}\left(\left(\mathbf{I}_{k}+\mathbf{D}\right)t_{s},\mathrm{diag}\left\{ \left(\mathbf{I}_{k}+\mathbf{D}\right)t_{s}\right\} \right),
\end{alignat*}
where $\mathbf{1}_{2c}\in\mathbb{R}^{1\times2c}$ is a row vector
of ones (hence $M$ is a matrix which sums the squared real and imaginary
parts of electric fields of all wavelngths), $\mathbf{u}$ is DM control
in WFE basis, $\cdot{}^{\circ2}$ stands for elementwise squaring,
and $\mathrm{diag}\left\{ \cdot\right\} $ yields a diagonal matrix
with the elements of its argument on the diagonal.

To avoid confusion with previous definitions, we denote EKF's covariance
(approximation) as $\hat{P}\in\mathbb{R}^{r\times r}$ and note that
it refers to \emph{open-loop} modes. It is advanced together with
the WFE estimate via (see \cite{stengel1994optimal})
\begin{alignat*}{1}
\hat{P}_{k+1|k} & =\hat{P}_{k|k}+Q,\\
\hat{P}_{k+1|k+1} & =\hat{P}_{k+1|k}-\hat{K}_{k+1}\hat{H}_{k+1}\hat{P}_{k+1|k},\\
\hat{\bm{\epsilon}}_{k+1|k} & =\hat{\bm{\epsilon}}_{k|k},\\
\hat{\bm{\epsilon}}_{k+1|k+1} & =\hat{\bm{\epsilon}}_{k+1|k}+\hat{K}_{k+1}\left(\mathbf{y}_{k+1}-\hat{\mathbf{y}}_{k+1}\right),
\end{alignat*}
with $\hat{K}_{k+1}\in\mathbb{R}^{r\times N_{pix}}$, $\hat{H}_{k+1}\in\mathbb{R}^{N_{pix}\times r}$
and $\hat{\mathbf{y}}_{k+1}\in\mathbb{R}^{N_{pix}}$ defined next.
The predicted photon count is given by
\[
\hat{\mathbf{y}}_{k+1}=\dot{N}_{S}t_{s}M\cdot\left(G\left(\hat{\bm{\epsilon}}_{k+1|k}+\mathbf{u}_{k+1}\right)+\mathbf{E}_{0}\right)^{\circ2}+\mathbf{D},
\]
its sensitivity to WFE is
\[
\hat{H}_{k+1}=\frac{\partial\hat{\mathbf{y}}_{k+1}}{\partial\hat{\bm{\epsilon}}_{k+1|k}}=2\dot{N}_{S}t_{s}M\mathrm{diag}\left\{ G\left(\hat{\bm{\epsilon}}_{k+1|k}+\mathbf{u}_{k+1}\right)+\mathbf{E}_{0}\right\} G,
\]
and the Kalman gain is
\[
\hat{K}_{k+1}=\hat{P}_{k+1|k}\hat{H}_{k+1}^{T}\left(\hat{H}_{k+1}\hat{P}_{k+1|k}\hat{H}_{k+1}^{T}+\mathrm{diag}\left\{ \hat{\mathbf{y}}_{k+1}\right\} \right)^{-1}.
\]

Finally, a control law $\mathbf{u}_{k+1}\left(\hat{\bm{\epsilon}}_{k+1|k}\right)$
must be provided. For Sec.~\ref{sub:RST} we sampled
\[
\mathbf{u}_{k+1}-\hat{\bm{\epsilon}}_{k+1|k}\sim{\cal N}\left(\mathbf{0},\sigma_{u}Q\right),
\]
where the dithering magnitude, $\sigma_{u}>1$, was chosen empirically
to give the best contrast.
\end{document}